\documentclass[12pt]{iopart}

\usepackage{epsfig}
\usepackage{graphicx}
\usepackage{float}
\usepackage{iopams}

\begin{document}

\title[Point processes in arbitrary dimension]{Point processes in arbitrary dimension from fermionic gases, random matrix theory,
and number theory}
\author{Salvatore Torquato$^1$, A Scardicchio$^2$ and Chase E Zachary$^3$}
\address{$^1$ Department of Chemistry, Princeton University, Princeton, New Jersey 08544, USA;\\
Program in Applied and Computational Mathematics, Princeton University, Princeton, New Jersey 08544, USA;\\
Princeton Institute for the Science and Technology of Materials, Princeton University, Princeton, New Jersey 08544, USA;\\
Princeton Center for Theoretical Science, Princeton University, Princeton, New Jersey 08544, USA;
and School of Natural Sciences, Institute for Advanced Study, Princeton, New Jersey, 08540, USA}
\ead{torquato@princeton.edu}
\address{$^2$ Department of Physics, Joseph Henry Laboratories, Princeton University,
Princeton, New Jersey 08544, USA; and
Princeton Center for Theoretical Science, Princeton University, Princeton, New Jersey 08544, USA}
\ead{ascardic@princeton.edu}
\address{$^3$ Department of Chemistry, Princeton University, Princeton, New Jersey 08544, USA}
\ead{czachary@princeton.edu}
%\date{\today}

\begin{abstract}

It is well known that one can map certain properties of random matrices, fermionic gases,  and zeros
of the Riemann zeta function to a unique point process on the real line $\mathbb{R}$.
Here we analytically provide exact  generalizations of such a point process in 
$d$-dimensional Euclidean space $\mathbb{R}^d$ for any $d$, which are
special cases of  determinantal processes. In particular, we obtain
the $n$-particle correlation functions for any $n$, which completely specify
the point processes in $\mathbb{R}^d$. We also demonstrate
that spin-polarized fermionic systems in $\mathbb{R}^d$ have these same $n$-particle
correlation functions in each dimension. The point processes
for any $d$ are shown to be hyperuniform, i.e., infinite wavelength density fluctuations
vanish, and the structure factor (or power spectrum) $S(k)$ has
a nonanalytic behavior at the origin given by $S(k) \sim |k|$ ($k \rightarrow 0$).
The latter result implies that the pair correlation function $g_2(r)$ 
tends to unity  for large pair distances with a decay rate
that is controlled by the power law $1/r^{d+1}$, which is a 
well-known property of bosonic ground states and more
recently has been shown to characterize maximally random jammed sphere packings.
We graphically display one- and two-dimensional realizations 
of the point processes in order to vividly reveal
their ``repulsive" nature. Indeed, we show that the point processes
can be characterized by an effective ``hard-core" diameter
that grows like the square root of $d$. The nearest-neighbor distribution functions 
for these point processes are also evaluated 
and rigorously bounded. Among other results, this analysis reveals 
that the probability of finding a large spherical cavity of radius $r$ in dimension $d$ behaves
like a Poisson point process but in dimension
$d+1$, i.e., this probability is given by $\exp[-\kappa(d) r^{d+1}]$ for large $r$ and finite $d$,
where $\kappa(d)$ is a positive $d$-dependent constant. We also show that as 
$d$ increases, the point process behaves effectively like  a sphere packing
with a coverage fraction of space that is no denser than $1/2^d$.
This coverage fraction has a special significance in the study of sphere 
packings in high-dimensional Euclidean spaces.

\end{abstract}

\pacs{02.50.Ey, 05.40.-a, 05.30.Fk}
\submitto{Journal of Statistical Mechanics:  Theory and Experiment}

\noindent{\it Keywords}: Correlation functions, Point processes, Fermions, Random matrix
theory, Number theory

\maketitle

\section{Introduction}

It is well known that there is a remarkable connection
between the statistical properties of certain random Hermitian matrices, the zeros of the Riemann zeta function,
and fermionic gases \cite{Dy62,Me67,Dy70,Mo73,Od87,Ru96}.  Underlying each of these objects are certain one-dimensional point processes (defined more precisely in Section 2)
whose statistical properties (under certain limits) are believed
to be identical. The purpose of this paper is to provide generalizations of
this unique point process to point processes in
$d$-dimensional Euclidean space $\mathbb{R}^d$ for any $d$,
and to characterize their spatial statistics exactly.
Since the characterization of a  point process can be viewed as
the study of a system of interacting ``point" particles,
exact descriptions of nontrivial point processes
in arbitrary space dimension, which are hard to come by \cite{To06b},
are of great value in the field of statistical mechanics.

There are three prominent theories of random Hermitian matrices, which
model the Hamiltonians of certain random dynamical systems; see
the excellent book by Mehta \cite{Me67}. If the dynamical system is symmetric under time reversal, then the relevant theory for integral spin is that of the Gaussian orthogonal ensemble (GOE) or the Gaussian symplectic ensemble (GSE) for half-integer spin. On the other hand, the Gaussian unitary ensemble (GUE) models random Hamiltonians without time reversal symmetry, which is relevant to certain properties of the Riemann zeta function.
 Although there are distinct one-dimensional point processes associated
with each of these ensembles, our interest here will be in the GUE because
of its relationship to the Riemann zeta function.

\begin{figure}[bthp]
\centerline{\psfig{file=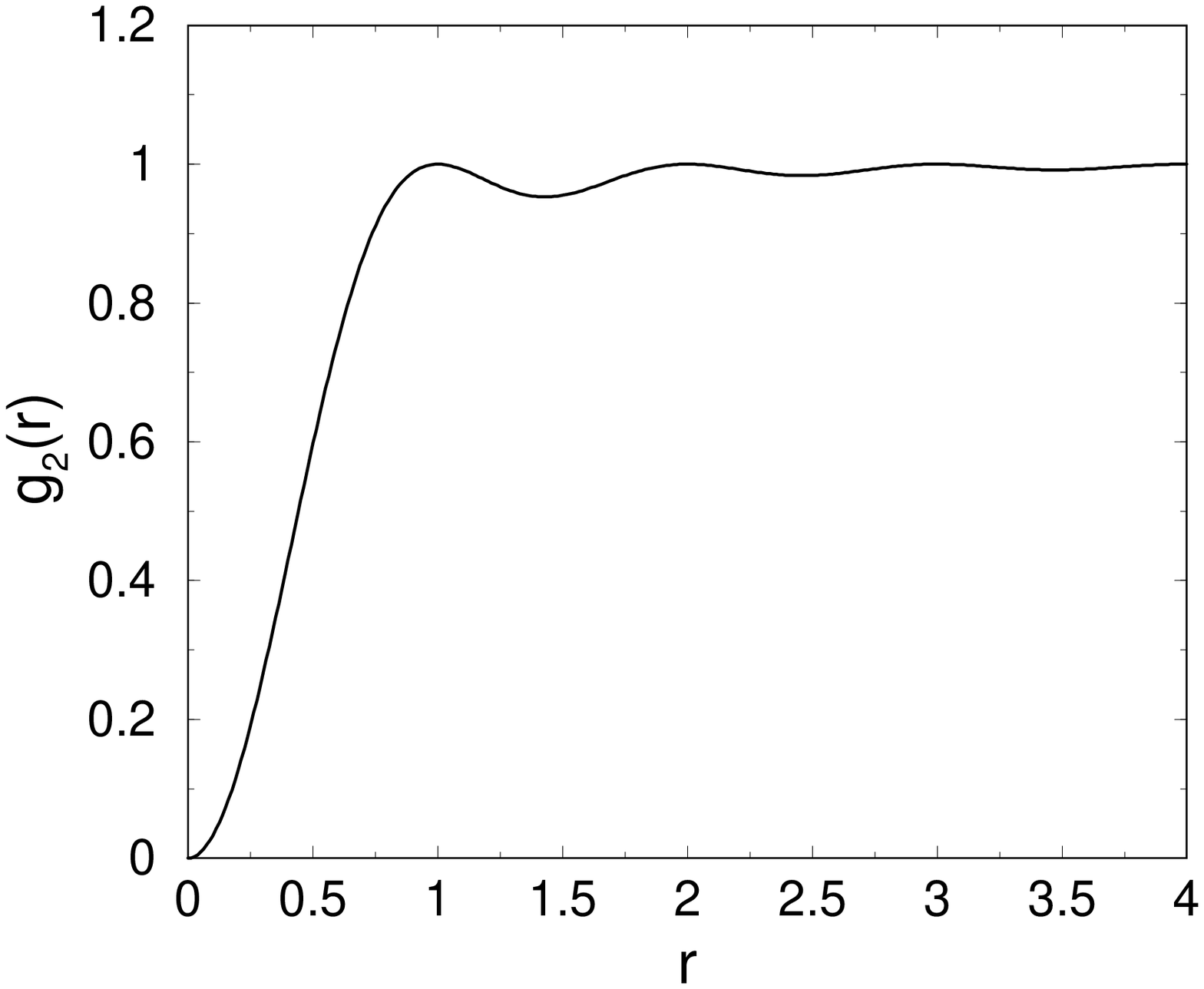,width=3.0in}\hspace{0.3in}\psfig{file=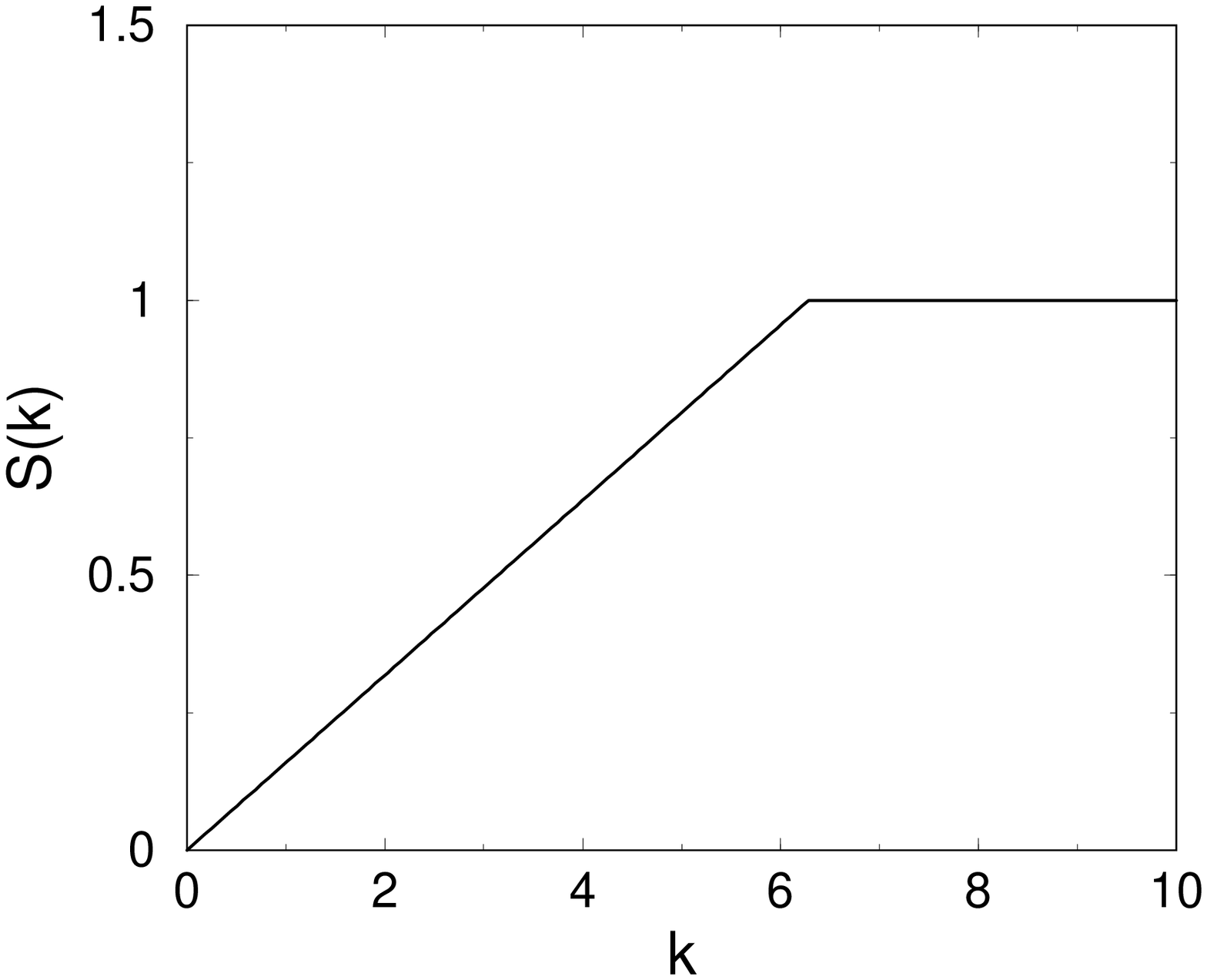,width=3.0in}}
\caption{Left panel: The pair correlation function $g_2(r)$ versus
distance $r$ for the eigenvalues of the GUE/CUE in the large-$N$ limit,
which is conjectured to be the same as the one characterizing the nontrivial zeros of the Riemann zeta function.
Spin-polarized fermions in their ground state in $\mathbb{R}$ have the same pair correlation function.
Right panel: The corresponding spectral counterpart, called the  structure factor $S(k)$ [cf. (\ref{factor})],
as a function of wavenumber $k$.}
\label{zeta-zero}
\end{figure}

The GUE of degree $N$ consists of the set of all $N\times N$ Hermitian matrices together with a Haar measure. 
This is the unique probability measure on the set of $N\times N$ Hermitian matrices that is invariant under 
conjugation by unitary matrices such that the individual matrix entries are independent random variables. 
Dyson \cite{Dy62} showed that the eigenvalue distrubutions
 of the GUE are closely related to those of the ``circular unitary ensemble" (CUE),
which he exactly mapped into a problem of point particles
on a unit circle interacting with a two-dimensional Coulombic force law
at a particular temperature. This point process on the unit circle  or, equivalently, 
on the real line $\mathbb{R}$ in the large-$N$ limit (when suitably normalized)
has a pair correlation function (defined in Section 2) in $\mathbb{R}$ at number density $\rho=1$  given by
\begin{eqnarray}
g_2(r)=1-\frac{\sin^2(\pi r)}{(\pi r)^2}.
\label{g2-zeros}
\end{eqnarray}  
Equation \eref{g2-zeros} also applies for the GUE in the limit $N\rightarrow +\infty$ such
that the mean gap distance between eigenvalues at the origin is normalized; this limit 
has the effect of magnifying the bulk of the eigenvalue density on $\mathbb{R}$ such that
\eref{g2-zeros} is well-defined.
We see that this point process is always {\it negatively correlated}, i.e., $g_2(r)$ never exceeds unity
(see Fig. \ref{zeta-zero}) and is ``repulsive" in the sense
that $g_2(r) \rightarrow 0$ as $r$ tends to zero. More generally, Dyson \cite{Dy70} proved that the $n$-particle correlation function (defined in Section 2) is given by the following
determinant:
\begin{eqnarray}
g_n(r_{12},r_{13},\ldots,r_{1n})=\mbox{det}\left( \frac{\sin(\pi r_{ij})}{\pi r_{ij}}\right)_{i,j=1,\ldots,n}.
\label{gn-zeros}
\end{eqnarray}

Montgomery \cite{Mo73} conjectured that the pair correlation function of nontrivial zeros of the Riemann zeta function
(which, according to the Riemann hypothesis, lie along the critical line $1/2 + it$ with $t\in\mathbb{R}$) 
is given exactly by the GUE/CUE function (\ref{g2-zeros}) in the asymptotic limit
(high on the critical line) when appropriately normalized. This remarkable correspondence
was further established by Odlyzko \cite{Od87}, who numerically verified the Riemann
hypothesis for the first $10^{13}$ nontrivial zeros of the zeta function as well
as at much larger heights and confirmed that the pair correlation function
agrees with (\ref{g2-zeros}). Rudnick and Sarnak \cite{Ru96}
proved that, under the Riemann hypothesis, the nontrivial zeros
have $n$-particle densities for any $n$ given by (\ref{gn-zeros}).
The reader is referred to the excellent review article
by Katz and Sarnak \cite{Ka99}, which discusses the connection
between the zeros of zeta functions and classical symmetric
groups, of which the three canonical random-matrix ensembles
are but special cases.

For spin-polarized free fermions in $\mathbb{R}$ (fermion gas) at number density $\rho=1$, it is known that the pair correlation function in the {\it ground state} (i.e., completely filling
the Fermi ``sphere") is given by
\begin{eqnarray}
g_2(r)=1-\frac{\sin^2(k_F r)}{(k_F r)^2},
\label{g2-fermion}
\end{eqnarray}  							    
where $k_F$ is the Fermi radius, which is the one-dimensional
analog of the well-known three-dimensional result \cite{Fe98}. 
Therefore, we see that when $k_F=\pi$, we obtain
the GUE/CUE pair correlation function (\ref{g2-zeros}). The repulsive nature
of the point process in this context arises physically from
the Pauli exclusion principle.

Very little is known about how to generalize one-dimensional point processes
associated with random matrices and number-theoretic functions to higher
dimensions. In this paper, we analytically obtain exact generalizations of
the aforementioned one-dimensional point process in 
$d$-dimensional Euclidean space $\mathbb{R}^d$ for any $d$. These processes are
special cases of  determinantal processes. 
In particular, we obtain
the $n$-particle correlation functions for any $n$, which completely specify
the point processes in $\mathbb{R}^d$. We also show
that spin-polarized fermionic systems in $\mathbb{R}^d$ have these same $n$-particle
correlation functions in each dimension. 
We show that the point processes for any $d$ are hyperuniform, i.e., infinite wavelength density fluctuations vanish, and the structure factor (or power spectrum) $S(k)$ has
a nonanalytic behavior at the origin given by $S(k) \sim |k|$ ($k \rightarrow 0$).
The latter result implies that the pair correlation function $g_2(r)$ 
tends to unity  for large pair distances with a decay rate
that is controlled by the power law $1/r^{d+1}$. 
In three dimensions,
such a dominant power-law decay of $g_2(r)$ is a well-known
property of bosonic systems in their ground states \cite{Fe54,Re67} and, more
recently, has been shown to characterize  maximally random jammed sphere packings \cite{Do05}.

Realizations of the point processes
are displayed in one and two dimensions, using a simulation technique described
by us elsewhere \cite{Sc08}, which vividly reveal their ``repulsive" nature.
In fact, we show the point processes
can be characterized by an effective ``hard-core" diameter
that grows like the square root of $d$.
The nearest-neighbor distribution functions 
for these point processes are also studied by evaluating 
and rigorously bounding them. Among other results, this analysis reveals that the probability of finding a 
large spherical cavity of radius $r$ in dimension $d$ behaves
like a Poisson point process but in dimension
$d+1$, i.e., this probability is given by $\exp[-\kappa(d) r^{d+1}]$ for large $r$,
where $\kappa(d)$ is a positive $d$-dependent constant. We also show that as 
$d$ increases, the point process behaves effectively like  a sphere packing
with a coverage fraction of space that is no denser than $1/2^d$.
As we will see, this value of the coverage fraction has a special
significance in the study of sphere packings in high dimensions.

In Section 2, we present background and definitions concerning point processes
that are particularly germane to the ensuing analysis.
Section 3 discusses general determinantal point processes.
In Section 4, we obtain the determinantal point processes
in $\mathbb{R}^d$ that are generalizations of the aforementioned
one-dimensional point process associated with fermions,
random matrices, and the Riemann zeta function. We call
the most general of these point processes ``Fermi-shells"
point processes. The asymptotic properties of various pair statistics
are investigated. We show
that spin-polarized fermionic systems in $\mathbb{R}^d$ have the same $n$-particle
correlation functions in each dimension. Section 5 analyzes
various nearest-neighbor functions for the special
case of the ``Fermi-sphere" point processes in $\mathbb{R}^d$.
We present concluding comments in Section 6, including remarks
about possible connections of our point processes to 
random matrix  and number theory.

\section{Background on Point Processes}
\subsection{Definitions}

A stochastic point process  in $\mathbb{R}^d$ 
is defined as a mapping from a probability space
to configurations of points ${\bf x}_1, {\bf x}_2, {\bf x}_3\ldots$
in $d$-dimensional Euclidean space $\mathbb{R}^d$. More precisely, let $X$ denote
the set of configurations such that each configuration 
$x \in X$ is a subset of $\mathbb{R}^d$ that satisfies two regularity
conditions: (i) there are no multiple points
(${\bf x}_i \neq {\bf x}_j$ if $ i\neq j$) and (ii)
each bounded subset of $\mathbb{R}^d$ must contain
only a finite number of points of $x$.
We denote by $N(B)$  the number of points within
$x \cap B$, $B \in {\cal B}$, where ${\cal B}$ is
the ring of bounded Borel sets in $\mathbb{R}^d$. Thus, we always
have $N(B) < \infty$ for $B \in {\cal B}$, but the
possibility $N(\mathbb{R}^d)=\infty$ is not excluded.
We note that there exists a minimal $\sigma$-algebra ${\cal U}$ of subsets
of $X$ that renders all of the functions $N(B)$ measurable.
Let $(\Omega,{\cal F}, {\cal P})$ be a probability
space. Any measurable map $x: \Omega \rightarrow X$
is called a stochastic point process \cite{St95,To06}.
Henceforth, we will simply call this map a point process.
Note that this random setting is quite general.
It incorporates cases in which the location of the points are deterministically
known, such as a lattice.

A point process is completely statistically characterized by specifying
the countably 
infinite set of $n$-particle probability density functions $\rho_n({\bf r}_1,{\bf r}_2,\ldots,{\bf r}_n)$ 
($n=1,2,3\ldots$) \cite{To06}.
The distribution-valued function $\rho_n({\bf r}_1,{\bf r}_2,\ldots,{\bf r}_n)$ 
has a probabilistic interpretation: apart from trivial constants,
it is the probability density function
associated with finding $n$ different points at positions $\mathbf{r}_1, \ldots, \mathbf{r}_n$ 
and hence has the nonnegativity property
\begin{eqnarray}
\rho_n({\bf r}_1,{\bf r}_2,\ldots,{\bf r}_n) \ge 0 \qquad \forall {\bf r}_i \in \mathbb{R}^d \quad (i=1,2,\ldots n).
\label{positive}
\end{eqnarray}
The point process is statistically homogeneous or translationally invariant if for every constant vector $\bf y$ in $\mathbb{R}^d$, 
$\rho_n({\bf r}_1,{\bf r}_2,\ldots,{\bf r}_n)=\rho_n({\bf r}_1+{\bf y},\ldots,{\bf r}_n+{\bf y})$,
which implies that   
\begin{eqnarray}
\rho_n({\bf r}_1,{\bf r}_2,\ldots,{\bf r}_n)=\rho^ng_n({\bf r}_{12},\ldots, {\bf r}_{1n}),
\label{nbody}
\end{eqnarray}
where $\rho$ is the {\it number density} (number of points per unit volume
in the infinite-volume limit)
and $g_n({\bf r}_{12},\ldots, {\bf r}_{1n})$ is the {\it $n$-particle correlation function},
which depends on the relative positions ${\bf r}_{12}, {\bf r}_{13}, \ldots$,
where ${\bf r}_{ij} \equiv {\bf r}_j -{\bf r}_i$ and we have chosen the origin to be at ${\bf r}_1$.
We call $g_2({\bf r})=g_2(-{\bf r})$ the pair correlation function.

For translationally invariant point processes without {\it long-range order}, 
$g_n({\bf r}_{12},\ldots, {\bf r}_{1n}) \rightarrow 1$ when
the points (or ``particles") are mutually far from one another, i.e.,  
as $|{\bf r}_{ij}| \rightarrow\infty$ 
($1\leq i < j < \infty$), $\rho_n({\bf r}_{1}, {\bf r}_2, \ldots, {\bf r}_{n}) \rightarrow \rho^n$.
Thus, the deviation of $g_n$ from unity  provides a
measure of the degree of spatial correlation
between the particles with unity corresponding to no spatial correlation.
Note that for a translationally invariant Poisson point process, 
$g_n$ is unity for all values of its argument. 

It is useful to introduce
the total correlation function $h({\bf r})$ of a translationally invariant point process, which
is related to the pair correlation function via
\begin{eqnarray}
h({\bf r})\equiv g_2({\bf r})-1
\label{total}
\end{eqnarray}
and decays to zero for large $|{\bf r}|$ in the absence of long-range order. 
Note that $h({\bf r})=0$ for
all $\bf r$ for a translationally invariant Poisson point process.
An important nonnegative spectral function $S({\bf k})$, called the structure factor (or power
spectrum), is defined
as follows: 
\begin{eqnarray}
S({\bf k}) =1+\rho{\tilde h}({\bf k}),
\label{factor}
\end{eqnarray}
where ${\tilde h}({\bf k})$ is the Fourier transform of $h(\bf r)$.  
For a translationally and rotationally invariant point process, it is
useful to consider the cumulative coordination number $Z(r)$, defined to be the expected
number of points found in a sphere of radius $r$ centered
at an arbitrary point of the point process, which is related to the
pair correlation function as follows:
\begin{eqnarray}
Z(r)=\rho s_1(1) \int_0^r x^{d-1} g_2(x) \rmd x,
\label{Z}
\end{eqnarray}
where
\begin{eqnarray}
s_1(r)  =  \frac{2\pi^{d/2}r^{d-1}}{\Gamma(d/2)}
\label{area-sph}
\end{eqnarray}
is the surface area of a  $d$-dimensional sphere of radius $r$.
It is clear that since $g_2(r)$ is a nonnegative function,
$Z(r)$ is a monotonically increasing function of $r$.

The Fourier transform of an $L^1$ function $f : \mathbb{R}^d \rightarrow \mathbb{R}$
is defined by
\begin{eqnarray}
{\tilde f}({{\bf k}})=\int_{\mathbb{R}^d} f({{\bf r}}) \exp\left[-\rmi({\bf k}\cdot{\bf r})\right] \rmd{{\bf r}},
\label{spec}
\end{eqnarray}
where $L^1$  denotes the space of absolutely integrable functions on $\mathbb{R}^d$.
If  $f : \mathbb{R}^d \rightarrow \mathbb{R}$ is a radial function, i.e., $f$ depends only
on the modulus $r=|\mathbf{r}|$ of the vector $\bf r$, then
its Fourier transform is given by
\begin{eqnarray}
{\tilde f}(k) =\left(2\pi\right)^{\frac{d}{2}}\int_{0}^{\infty}r^{d-1}f(r)
\frac{J_{\left(d/2\right)-1}\!\left(kr\right)}{\left(kr\right)^{\left(d/2\right
)-1}} \,\rmd r,
\label{fourier}
\end{eqnarray}
where $k$ is the modulus of the wave vector $\bf k$
and $J_{\nu}(x)$ is the Bessel function of order $\nu$.
The inverse transform of $\tilde{f}(k)$ is given by
\begin{eqnarray}
f(r) =\frac{1}{\left(2\pi\right)^{\frac{d}{2}}}\int_{0}^{\infty}k^{d-1}\tilde{f}(k)
\frac{J_{\left(d/2\right)-1}\!\left(kr\right)}{\left(kr\right)^{\left(d/2\right
)-1}}\rmd k.
\label{inverse}
\end{eqnarray}

\subsection{Number Variance and Hyperuniformity}

We denote by $\sigma^2(A)$ 
the variance in the number of points $N(A)$ contained within a window
$A \subset \mathbb{R}^d$. The number variance $\sigma^2(A)$ for a specific choice
of $A$ is necessarily a positive number and is
generally related to the total  correlation function $h(\bf r)$ for
a translationally invariant point process \cite{To03a}.
In the special case of a spherical window of radius $R$ in $\mathbb{R}^d$,
it is explicitly given by
\begin{eqnarray}
\sigma^2(R)=\rho v_1(R) \Bigg[ 1+\rho \int_{\mathbb{R}^d} h({\bf r}) \alpha(r; R) \, \rmd{\bf r}\Bigg], 
\label{variance}
\end{eqnarray}
where $\sigma^2(R)$ is the number variance for
a spherical window of radius $R$ in $\mathbb{R}^d$
and $\alpha(r;R)$ is the volume common to two spherical windows of radius $R$
whose centers are separated by a distance $r$ divided by $v_1(R)$, 
the  volume of a spherical window of radius $R$, given explicitly by
\begin{eqnarray}
v_1(R)= \frac{\pi^{d/2}}{\Gamma(1+d/2)}R^d.
\label{v1}
\end{eqnarray}
We will call $\alpha(r;R)$ the {\it scaled intersection volume}.

For large $R$, it has been proved that $\sigma^2(R)$ cannot grow more
slowly than $\gamma R^{d-1}$, where $\gamma$ is a positive constant \cite{Beck87}.
We note that point processes (translationally invariant or not)
for which $\sigma^2(R)$ grows more slowly than the window volume (i.e.,  
as  $R^{d}$) for large $R$ are examples
of {\it hyperuniform} (or superhomogeneous) point patterns \cite{To03a,Ga03}. This classification includes all periodic point processes \cite{To03a}, 
certain aperiodic point processes \cite{Ga03,To03a}, one-component plasmas \cite{Ga03,To03a},
point processes associated with a wide class of  tilings of space \cite{Ga04,Ga08},
and certain disordered sphere packings \cite{To06,To02c,Sc08b,footnote}. Hyperuniformity
implies that the structure factor $S({\bf k})$ has the following
small ${\bf k}$ behavior:
\begin{eqnarray}
\lim_{{\bf k}\rightarrow {\bf 0}} S({\bf k}) =0.
\label{hyper}
\end{eqnarray}

\begin{figure}[bthp]
\centerline{\psfig{file=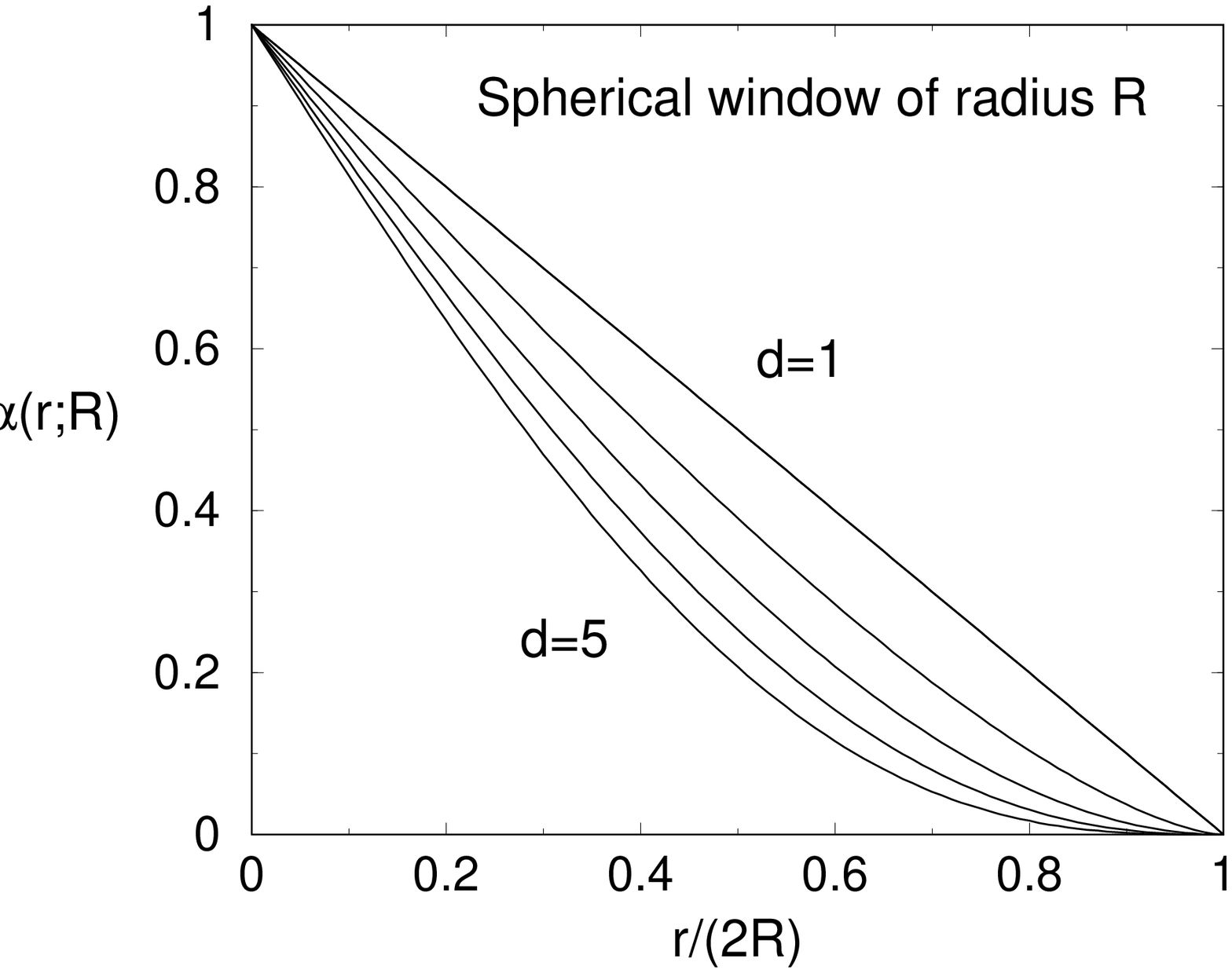,width=2.8in} \hspace{0.2in} \psfig{file=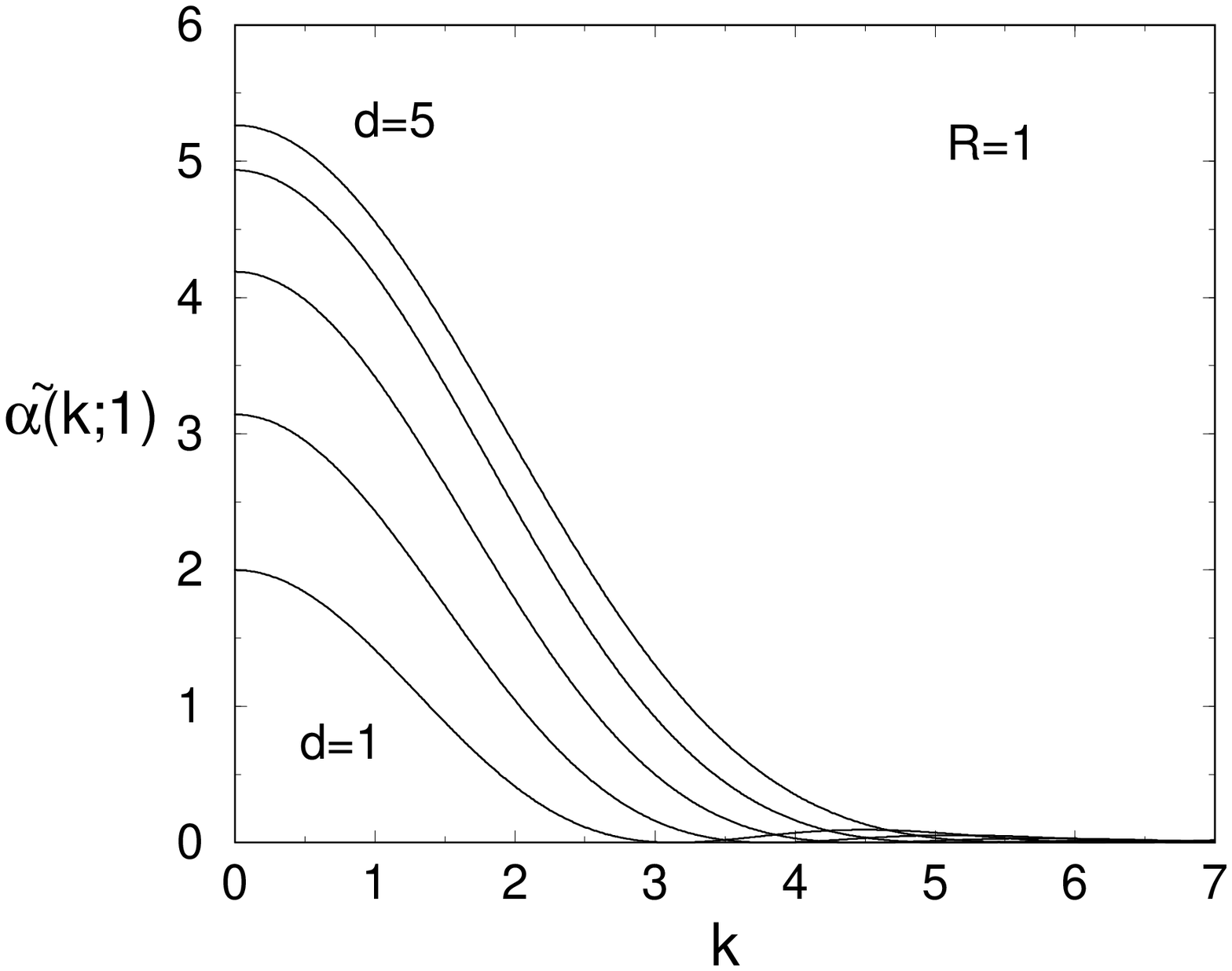,width=3.0in}}
\caption{Left panel: The scaled intersection
volume $\alpha(r;R)$ for spherical windows of radius $R$
as a function of $r$ for the first
five space dimensions. The uppermost curve is for $d=1$
and lowermost curve is for $d=5$. Right panel: Corresponding Fourier transforms
for the case $R=1$. The uppermost curve is for $d=5$
and lowermost curve is for $d=1$.}
\label{intersection}
\end{figure}

The scaled intersection volume  $\alpha(r;R)$ appearing in (\ref{variance}) and
its associated Fourier transform will play a prominent role
in this paper. The former quantity is defined by a convolution integral involving the  indicator function $w$
for a $d$-dimensional spherical ``window" of radius $R$ centered
at position ${\bf x}_0$ \cite{To03a}, i.e., 
\begin{eqnarray}
w(|{\bf x}-{\bf x}_0|;R)=\Theta(R-|{\bf x}-{\bf x}_0|),
\label{window2}
\end{eqnarray}
where $\Theta(x)$ is the Heaviside step function
\begin{eqnarray}
%\Theta(x) =\Bigg\{{0, \quad x<0,\atop{1, \quad x \ge 0.}}\
\Theta(x) = \left\{
\begin{array}{lr}
0, \quad x<0,\\
1, \quad x\ge0.
\end{array}\right.
\label{heaviside}
\end{eqnarray}
Specifically, the scaled intersection volume is given by
\begin{eqnarray}\label{alpha1}
\alpha(r;R)= \frac{1}{v_1(R)}\int_{\mathbb{R}^d} w({\bf r}_1-{\bf x}_0;{\bf R})
w({\bf r}_2-{\bf x}_0;{\bf R}) \rmd{\bf x}_0.
\end{eqnarray}
The scaled intersection volume  has the support
$[0,2R]$, the range $[0,1]$, and the following alternative integral representation \cite{To06}:
\begin{eqnarray}
\alpha(r;R) = c(d) \int_0^{\cos^{-1}[r/(2R)]} \sin^d(\theta) \, \rmd\theta,
\label{alpha}
\end{eqnarray}
where $c(d)$ is the $d$-dimensional constant given by
\begin{eqnarray}
c(d)= \frac{2 \Gamma(1+d/2)}{\pi^{1/2} \Gamma[(d+1)/2]}.
\label{C}
\end{eqnarray}
Torquato and Stillinger \cite{To06} found the following  series representation
of the scaled intersection volume $\alpha(r;R)$ for $r \le 2R$
and for any $d$:
\begin{eqnarray}
\fl \alpha(r;R)=1- c(d) x+ 
c(d) \sum_{n=2}^{\infty}
(-1)^n \frac{(d-1)(d-3) \cdots (d-2n+3)}
{(2n-1)[2 \cdot 4 \cdot 6 \cdots (2n-2)]} x^{2n-1},
\label{series}
\end{eqnarray}
where $x=r/(2R)$. For even dimensions, relation (\ref{series}) is
an infinite series, but for odd dimensions, the series truncates such
that $\alpha(r;R)$ is a univariate polynomial of degree $d$. 
In even dimensions, the scaled intersection volume involves
transcendental functions (e.g., for $d=2$,
$\alpha(r;R) =   2{\pi}^{-1} [ \cos^{-1}(\frac{r}{2R}) - \frac{r}{2R}
(1 - \frac{r^2}{4R^2})^{1/2} ]$ for $0 \le r \le 2R$).
Figure \ref{intersection} provides plots of $\alpha(r;R)$ as a function of $r$ for the first
five space dimensions. For any dimension, $\alpha(r;R)$
is a monotonically decreasing function of $r$. At a fixed
value of $r$ in the interval $(0,2R)$, $\alpha(r;R)$
is a monotonically decreasing function of the dimension $d$.

The Fourier transform of the scaled intersection volume function (\ref{alpha}),
which is given by
\begin{eqnarray}
{\tilde \alpha}(k;R) = \frac{[{\tilde w(k;R)}]^2}{v_1(R)},
\end{eqnarray}
where 
\begin{eqnarray} 
{\tilde w}(k;R) &= \frac{(2\pi)^{d/2}}{k^{(d/2)-1}}\int_0^R
r^{d/2} J_{(d/2)-1}(kr) \rmd r \nonumber \\
&= \left(\frac{2\pi}{kR}\right)^{d/2} R^d J_{d/2}(kR),
\label{window-Fourier}
\end{eqnarray}
is the Fourier transform of the window indicator function (\ref{window2}).
Therefore, the Fourier transform of $\alpha(r;R)$ is the following
nonnegative function of $k$:
\begin{eqnarray}
{\tilde \alpha}(k;R) = 2^d \pi^{d/2} \Gamma(1+d/2)\left(\frac{J_{d/2}(kR)}{k^{d/2}}\right)^2.
\label{alpha-k}
\end{eqnarray}

It has been shown that finding the point process that minimizes the number variance
$\sigma^2(R)$ is equivalent to finding the ground state of a certain repulsive pair potential
with compact support \cite{To03a}. This problem is directly related 
to an outstanding problem in number theory involving generalized zeta functions
and lattices \cite{Sa06}. Understanding such ground states can
be facilitated by utilizing ``duality" relations that link
ground states in real space to those in Fourier space \cite{To08}.

\section{Determinantal Point Processes}

We will be able to obtain the appropriate $d$-dimensional generalizations
of point processes corresponding to the eigenvalues of the GUE,
the zeros of the Riemann zeta function, or the one-dimensional fermionic gas
by appealing to the notion of a determinantal point process in $\mathbb{R}^d$ \cite{Ma75}.
Determinantal point processes were introduced
by Macchi \cite{Ma75}, who originally called them fermion point processes.
Soshnikov \cite{So00} presented a review of this subject and discussed
applications to random matrix theory, statistical mechanics,
quantum mechanics, and representation theory. It is also noteworthy
that examples of determinantal point processes arise
in self-avoiding random walks \cite{Jo04} and uniform spanning trees \cite{Bu93}.

Without loss of generality, the number density is set to unity ($\rho=1$) in
the ensuing discussion.
Let $H({\bf r})=H(-{\bf r})$ be  a translationally invariant Hermitian-symmetric kernel of an integral
operator $\cal H$. A translationally invariant {\it determinantal point process} in $\mathbb{R}^d$ exists if the
the $n$-particle density functions for $n \ge 1$ are given by the following
determinants:
\begin{eqnarray}
\rho_n({\bf r}_{12},{\bf r}_{13},\ldots,{\bf r}_{1n})=\mbox{det}[H({\bf r}_{ij})]_{i,j=1,\ldots,n},
\label{rhon-det}
\end{eqnarray}
where $H({\bf 0})=1$. By virtue of the nonnegativity of the $\rho_n$ in the pointwise sense [cf. (\ref{positive})]
and (\ref{rhon-det}), it follows that $\cal H$ must have nonnegative minors and $\cal H$ must
be a nonnegative operator, which implies that $H({\bf r})$ is positive semidefinite.
The latter implies that the  Fourier transform ${\tilde H}({\bf k})$ of the kernel $H({\bf r})$ is nonnegative,
and this property together with the condition $H({\bf 0})=1=\int_{\mathbb{R}^d} {\tilde H}({\bf k}) d{\bf k}$ 
implies that ${\tilde H}({\bf k}) \le 1$, i.e.,
\begin{eqnarray}
0 \le {\tilde H}({\bf k}) \le 1 \quad \mbox{for all} \; {\bf k}.
\label{f-K}
\end{eqnarray}
It follows that any  positive semidefinite Hermitian-symmetric kernel  
$H({\bf r})=H(-{\bf r})$ 
whose Fourier transform satisfies the inequalities in (\ref{f-K}) describes a determinantal
point process with a pair correlation
function given by
\begin{eqnarray}
g_2({\bf r})=1-|H({\bf r})|^2,
\label{g2-determ}
\end{eqnarray} 
such that
\begin{eqnarray}
0 \le g_2({\bf r}) \le 1 \qquad \mbox{and} \qquad g_2({\bf 0})=0,
\label{conditions}
\end{eqnarray}
and a $n$-particle density given by (\ref{rhon-det}).
We see that the total correlation function for a determinantal
point process is given by
\begin{eqnarray}
h({\bf r})=-|H({\bf r})|^2.
\label{h}
\end{eqnarray}

The fact that the $n$-particle density functions can be written
in terms of the determinant specified by (\ref{rhon-det}) leads
to bounds on $\rho_n$. For example, it trivially follows that
\begin{eqnarray}\label{rhotriv}
\rho_n({\bf r}_{12},{\bf r}_{13},\ldots,{\bf r}_{1n}) \le 1.
\end{eqnarray}
A less obvious but stronger upper bound is as follows:
\begin{eqnarray}
\rho_n({\bf r}_{12},{\bf r}_{13},\ldots,{\bf r}_{1n}) \le  \rho_2({\bf r}_{12})\rho_2({\bf r}_{13}) \cdots 
\rho_2({\bf r}_{1n}).
\label{rhon-ineq}
\end{eqnarray}
We remark also that:
\begin{eqnarray}\label{rhomn}
\rho_n(\mathbf{r}_{12},\ldots,\mathbf{r}_{1n}) \leq \rho_m(\mathbf{r}_{12},\ldots,\mathbf{r}_{1m})~~
\forall m\leq n.
\end{eqnarray}
Each of these inequalities is a consequence of the determinantal form for $\rho_n$
and the characteristics of $H$; 
in particular, \eref{rhomn} follows directly from Fischer's inequality \cite{HoJo05} and an 
appropriate partition of
the matrix representation for the operator $\mathcal{H}$.  
Equation \eref{rhon-ineq} is a result of the more general
Hadamard-Fischer inequalities \cite{HoJo05} and the relation $H(\mathbf{0}) = 1$.  
The positive semidefinite character of $\mathcal{H}$ is essential for these inequalities to hold.

A unique property of determinantal point processes is that each of the $n$-particle 
correlation functions $g_n$ can be expressed completely in terms of the pair correlation
function $g_2$.
Namely, at unit density we may write:
\begin{eqnarray}
g_2(r) = 1-\left[H(r)\right]^2\\
\Rightarrow H(r) = \pm\sqrt{1-g_2(r)}\label{Hsqrt}.
\end{eqnarray}
Therefore:
\begin{eqnarray}
g_n(\mathbf{r}_{12},\ldots,\mathbf{r}_{1n}) &= 
\det\left[H(r_{ij})\right]_{1\leq i<j\leq n}\nonumber\\
 &= \det\left[\pm\sqrt{1-g_2(r_{ij})}\right]_{1\leq i<j\leq n}\label{gng2}.
 \end{eqnarray}
The right side of
\eref{Hsqrt} is well-defined for all $r \in \mathbb{R}^+$ since $0\leq g_2(r) \leq 1~\forall r$.
We note that in general $H(r)$ may be either positive or negative for a given value of $r$
as in \eref{gn-zeros}; therefore, the sign of the square root in \eref{Hsqrt} must be 
chosen appropriately.
Our ability to express each $g_n$ in terms of the pair correlation function $g_2$ 
is a reflection of the fact that the $n$-particle correlation functions depend on a 
common kernel $H$; such a reformulation is generally not possible for an arbitrary point process.
Thus, one can infer the behavior of the $n$-particle correlation functions for
$n\geq 3$ solely from a knowledge of the behavior of $g_2$.

A trivial example
of a determinantal point process is the case in which $H(0)=1$ and $H(r)=0$ for  $r\neq 0$.
The resulting pair correlation function is given by $g_2(0) = 0$ and $g_2(r) = 1~\forall r\neq 0$,
and this function belongs to the same equivalence class as the pair correlation function of the Poisson
point process with respect to Lebesgue measure on the nonnegative reals $\mathbb{R}^+$
(i.e., the functions differ only on a set of measure zero).
Note that Costin and Lebowitz \cite{Cos04} have 
considered the conditions under which a pair correlation function
of the form $g_2(r)=1-\exp(-\lambda r)$ is a determinantal point process in $\mathbb{R}^d$.

We note in passing that the number of points in a determinantal point process that falls in
a compact set $A \subset \mathbb{R}^d$ has the same distribution as a sum
of independent Bernoulli($\lambda_i^A$) random variables, where $\lambda_i^A$
are the eigenvalues of the operator $\cal H$ \cite{Ho06}. Moreover, 
Hough et al. \cite{Ho06} presented an algorithm to generate determinantal
point process in $\mathbb{R}^d$, which we apply elsewhere \cite{Sc08}.

\section{New Determinantal Point Processes in $\mathbb{R}^d$ and Their Connection
to Fermionic Gases}
\label{determ-d}

\subsection{``Fermi-Sphere" Point Processes in $\mathbb{R}^d$}

Here we obtain the appropriate generalization of (\ref{g2-zeros})
that corresponds to a determinantal point process in $\mathbb{R}^d$.
First, we make the simple observation that the pair correlation function specified by (\ref{g2-zeros})
is related to the Fourier transform ${\tilde \alpha}(k;R)$ of the one-dimensional scaled intersection
volume evaluated at $k=r$ and $R=\pi$, namely,
\begin{eqnarray}
g_2(r)=1 - \frac{{\tilde \alpha}(r;\pi)}{2\pi},
\label{g2-one}
\end{eqnarray} 
where ${\tilde \alpha}(k;R)$ is given by (\ref{alpha-k}) for $d=1$.
A natural generalization of this pair correlation function in $\mathbb{R}^d$ is to replace the
one-dimensional Fourier transform in (\ref{g2-one})  
with its $d$-dimensional counterpart (\ref{alpha-k}) evaluated at $k=r$ 
and divided by $(2\pi)^d$, i.e.,
\begin{eqnarray}
g_2(r)=1 - \frac{{\tilde \alpha}(r;R)}{(2\pi)^d}.
\label{g2-d}
\end{eqnarray}
However, the value of $R$ in each dimension must be chosen so that
such a pair correlation function corresponds to a determinantal
point process in that dimension. In other words, if we take the positive semidefinite function
$H({\bf r})$ to be given by the following radial function:
\begin{eqnarray}
H(r)=\frac{\sqrt{{\tilde \alpha}(r;R)}}{(2\pi)^{d/2}}= \frac{\sqrt{\Gamma(1+d/2)}}{\pi^{d/4}}
\frac{J_{d/2}(rR)}{r^{d/2}},
\label{H}
\end{eqnarray}
$R$ must be determined so that the conditions (\ref{f-K}) and
(\ref{conditions}) are satisfied. Noting that the expansion of $|H(r)|^2$
for small $r$ to leading order is given by
\begin{eqnarray}
|H(r)|^2= \frac{R^d}{2^d \pi^{d/2} \Gamma(1+d/2)}  +\; {\cal O}(r^2)
\end{eqnarray}
and using the condition that $g_2(0)=0$ [cf. (\ref{conditions})] yields that 
\begin{eqnarray}
R=K \equiv 2 \sqrt{\pi}\, [\Gamma(1+d/2)]^{1/d}.
\label{R}
\end{eqnarray}

Now we must show that a pair correlation function (\ref{g2-d}) with $R=K$ satisfies
the bounds of (\ref{conditions}) and the bounds (\ref{f-K}) on
the spectral function ${\tilde H}(k)$. The square of the function
$H(r)$, specified by (\ref{H}), is clearly positive and achieves its maximum value
of unity at the origin when $R=K$ and tends to zero in the limit $r \rightarrow \infty$, 
and hence the bounds of (\ref{conditions}) are satisfied when $R=K$.
Referring to relation (\ref{window-Fourier}) for the Fourier transform
of the window indicator function, we see that the Fourier
transform of (\ref{H}) with $R=K$ is simply the indicator function
\begin{eqnarray}
{\tilde H}(k)=w(k;K)=\Theta(K-k),
\end{eqnarray}
which automatically satisfies (\ref{f-K}), where $w$ is specified by (\ref{window2}) and $K$ 
is given by (\ref{R}).

In summary, we have demonstrated that there is a determinantal point process in $\mathbb{R}^d$ 
with $n$-particle densities given by (\ref{rhon-det}) with the kernel
\begin{eqnarray}
H(r)=\frac{\sqrt{{\tilde \alpha}(r;K)}}{(2\pi)^{d/2}},
\label{H2}
\end{eqnarray}
where $K$ is given by (\ref{R}).
We will call such a determinantal point process a ``Fermi-sphere" point process
because ${\tilde H}(k)$ corresponds to a spherical window
indicator function in Fourier space and, as we will see, corresponds
to the ground state of a fermionic system in which the Fermi sphere 
is completely filled (see Section 3). 
In particular, the pair correlation function of such a point process \cite{anydensity}
is given by
\begin{eqnarray}
g_2(r)= 1-2^d\Gamma(1+d/2)^2\frac{J^2_{d/2}(Kr)}{(Kr)^{d}}.
\label{g2-d-2}
\end{eqnarray}
Moreover, the corresponding structure factor, at unit number density, takes the form
\begin{eqnarray}
S(k)=1-\alpha(k;K),
\label{S-d}
\end{eqnarray}
where $\alpha(k;K)$ is the scaled intersection volume of two $d$-dimensional spheres
of radius $K$ separated by a distance $k$, i.e., the function $\alpha(r;R)$,
specified by (\ref{series}), with the replacements $r \rightarrow k$ and $R \rightarrow K$.
Relation (\ref{S-d}) is easily obtained by taking the Fourier
transform of the total correlation function defined by (\ref{h}), where $H(r)$ is given by 
(\ref{H2}), and employing the definition (\ref{factor}) for the structure factor.
It follows from the properties of $\alpha(k;K)$  in (\ref{S-d})
that the structure factor $S(k)$ obeys the following bounds for all $k$:
\begin{eqnarray}
0 \le S(k) \le 1,
\end{eqnarray}
and  achieves its maximum value of unity for $k \ge 2K$.
The corresponding cumulative coordination number is given by
\begin{eqnarray}
Z(r)= v_1(r) - d \int_0^r \frac{J^2_{d/2}(Kx)}{x} \rmd x.
\label{Z-2}
\end{eqnarray}
Note that the first term in (\ref{Z-2}) is precisely
the cumulative coordination number for a Poisson point
process in $\mathbb{R}^d$ and that the second term 
is strictly negative for any $r>0$, which is a reflection
of the short-range repulsive nature of the point process. We will show that for
sufficiently large $r$, the dominant contribution to
$Z(r)$ will be the Poissonian term $v(r)$. We will see
it is the cumulative coordination number $Z(r)$, rather
than the pair correlation function (contrary to conventional
wisdom), that enables
one to identify and quantify an effective
``hard-core" diameter of the determinantal point processes.

It is instructive to examine the asymptotic behaviors of  $g_2(r)$, $S(k)$, and $Z(r)$
for the Fermi-sphere point process.
By virtue of the asymptotic properties of the Bessel function of arbitrary order,
the small-$r$ and large-$r$ forms of the pair correlation function (\ref{g2-d-2}) are respectively given by
\begin{eqnarray}
g_2(r)= \frac{K^2}{d+2} r^2 - \frac{(d+3)K^4}{2(d+2)^2(d+4)} r^4 +\; {\cal O}(r^6)  \qquad (r \rightarrow 0)
\label{quad}
\end{eqnarray}
and
\begin{eqnarray}
g_2(r)=1 - \frac{2\Gamma(1+d/2)\cos^{2}\,(rK -\pi(d+1)/4)}{K\, \pi^{d/2+1}\, r^{d+1}} \qquad (r \rightarrow \infty).
\end{eqnarray}
We see that $g_2(r)$ tends to zero quadratically in $r$ in the limit $r \rightarrow 0$,
independent of the dimension. The coefficient of the quadratic
term in (\ref{quad}) for positive $d$ attains it maximum
value of $\pi^2/3=3.2898\ldots$ for $d=1$ and monotonically decreases
with increasing dimension, asymptoting to $2\pi/e=2.3114\ldots$
in the limit $d\rightarrow\infty$. Moreover, $g_2(r)$ tends to unity  for large pair distances 
with a decay rate that is controlled by the power law $1/r^{d+1}$
for any $d \ge 1$. 

The latter result implies 
that the structure factor $S(k)$ tends to zero linearly in $k$ in
the limit $k\rightarrow 0$. Using (\ref{series}) and (\ref{S-d}), it is easy
to verify that as $k$ tends to zero at $\rho=1$
\begin{eqnarray}
S(k)=\frac{c(d)}{2K}k +\; {\cal O}(k^3)  \qquad (k \rightarrow 0),
\label{linear}
\end{eqnarray}
where $c(d)$ is a $d$-dependent positive constant given by (\ref{C}).
We see that at unit density,  the point process
in $\mathbb{R}^d$ for any $d$ is hyperuniform, and the structure factor $S(k)$ has
a nonanalytic behavior at the origin given by $S(k) \sim |k|$ ($k \rightarrow 0$).
 The coefficient of the linear term in (\ref{linear}) for positive $d$ attains it minimum
value of $1/(2\pi)=0.1591\ldots$ for $d=1$ and monotonically increases
with increasing dimension such that it asymptotes to $e^{1/2}/(2\pi)=0.2624\ldots$
in the limit $d\rightarrow\infty$. In three dimensions, this unusual linear nonanalytic behavior
of the structure factor at $k=0$ is a well-known
property of bosons in their grounds states \cite{Fe54,Fe56,Re67} and, more recently, has been 
shown to characterize maximally random jammed sphere packings \cite{Do05}.

\begin{figure}[bthp]
\centerline{\psfig{file=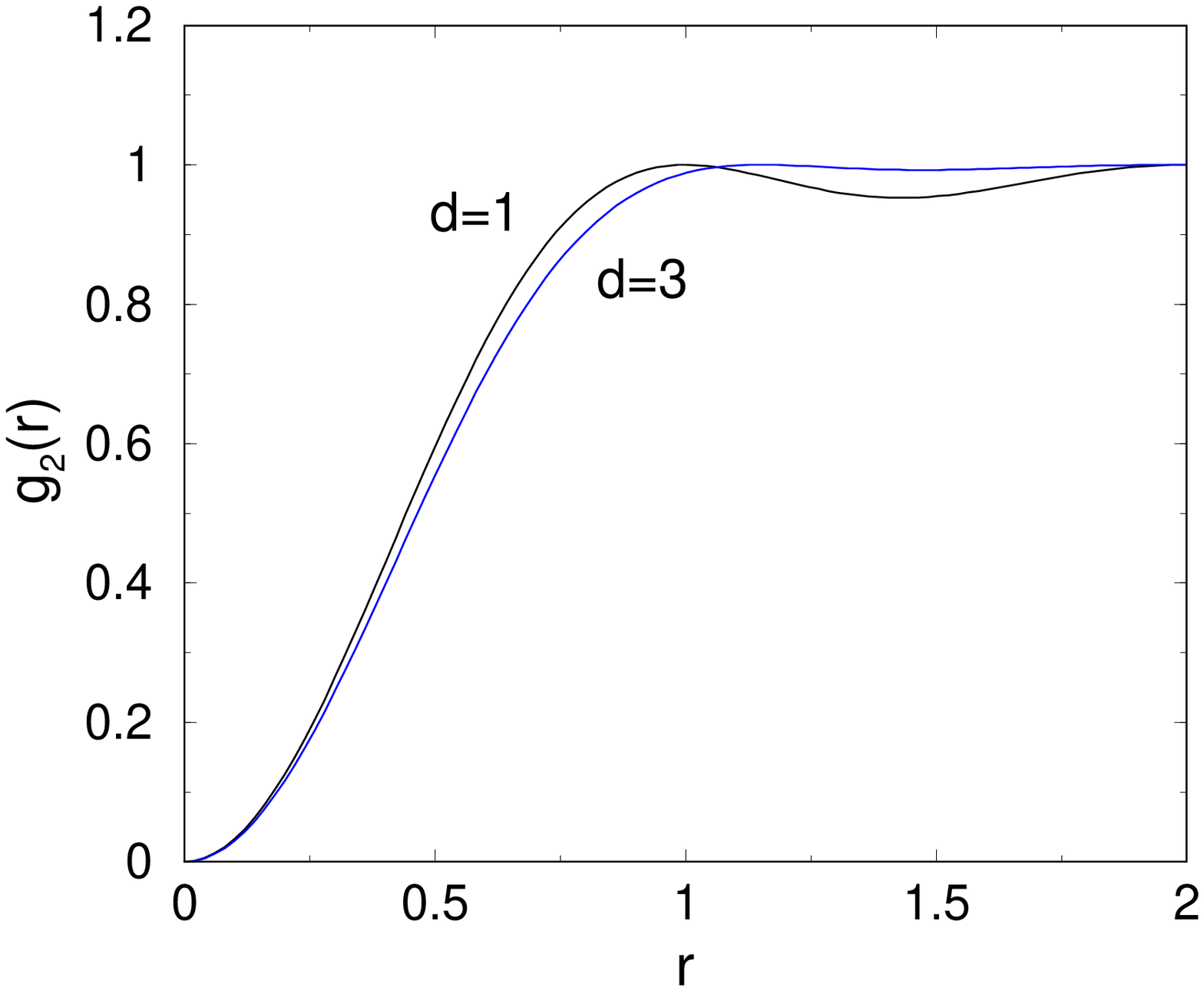,width=3.0in}\hspace{0.3in}\psfig{file=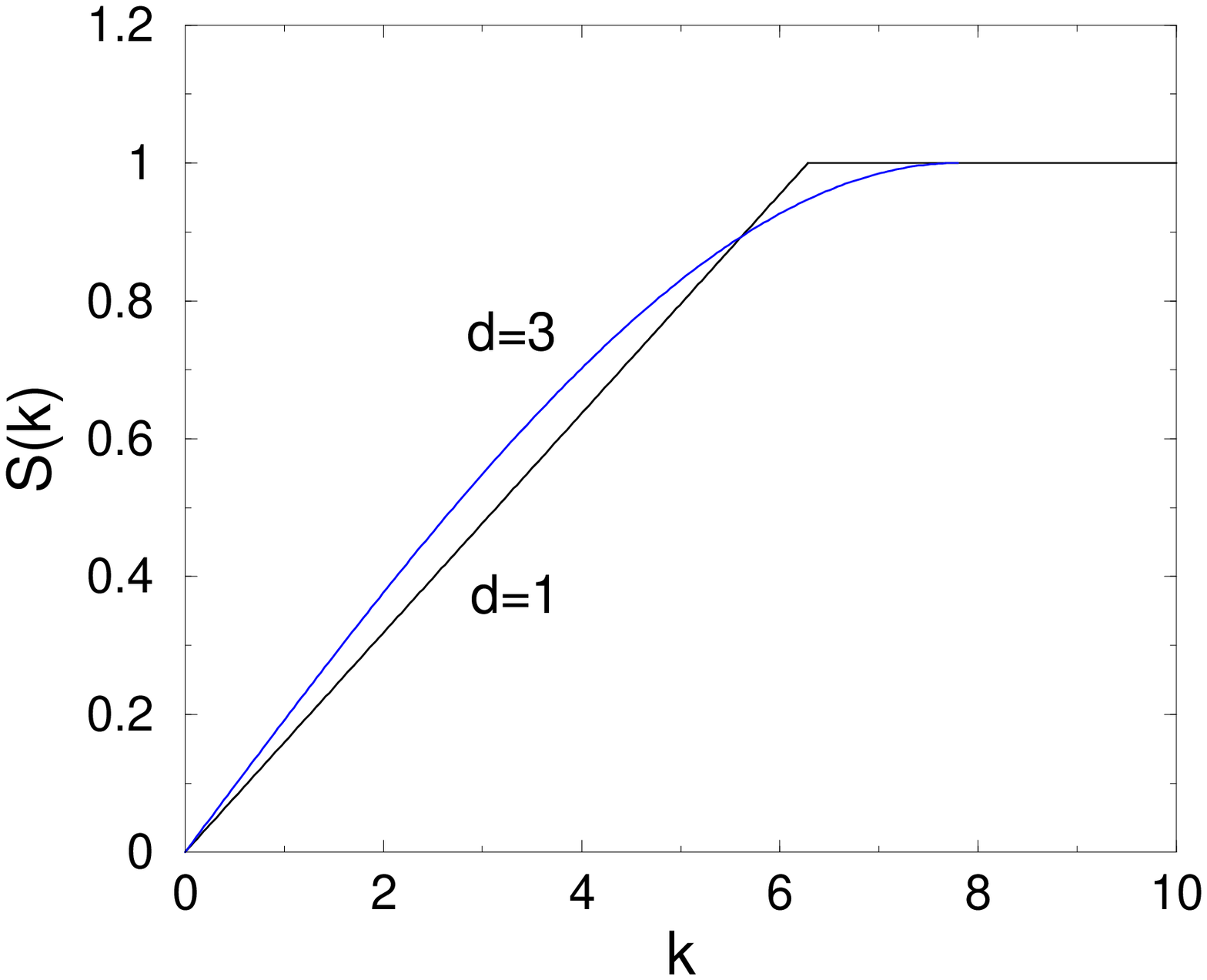,width=3.0in}}
\caption{Left panel: Comparison of the pair correlation functions for $d=1$ and $d=3$.
Right panel: The corresponding structure factors.}
\label{fermions}
\end{figure}

In Fig. \ref{fermions}, we compare pair statistics
in both real and Fourier space for $d=1$ and $d=3$
at unit density. We see that the amplitudes of the oscillations
in $g_2(r)$ that are apparent for $d=1$ are significantly 
reduced in the corresponding three-dimensional pair correlation 
function. The smallest value of $r$ for which $g_2(r)$ attains
its maximum value of unity, which we denote by $r_0$, is determined by the first 
positive zero of $J_{d/2}(Kr)$, which for sufficiently
large $d$ is given by
\begin{eqnarray}
\fl Kr_0 = \frac{d}{2} + 1.4729154 d^{\,1/3} + \frac{1.301687}{d^{\,1/3}} 
- \frac{0.007942}{d}+{\cal O}\left(\frac{1}{d^{\,5/3}}\right) \qquad (d \rightarrow \infty).
\end{eqnarray} 
Since $K$ increases with increasing $d$ and grows
like $\sqrt{2\pi d/e}$ for large $d$,  $r_0$ grows
like the square root of $d$ for large $d$. Similarly, the smallest value
of $k$ for which $S(k)$ attains its maximum value
of unity grows with increasing $d$ and for large $d$ grows
like $\sqrt{d}$.

\begin{figure}[bthp]
\centerline{\psfig{file=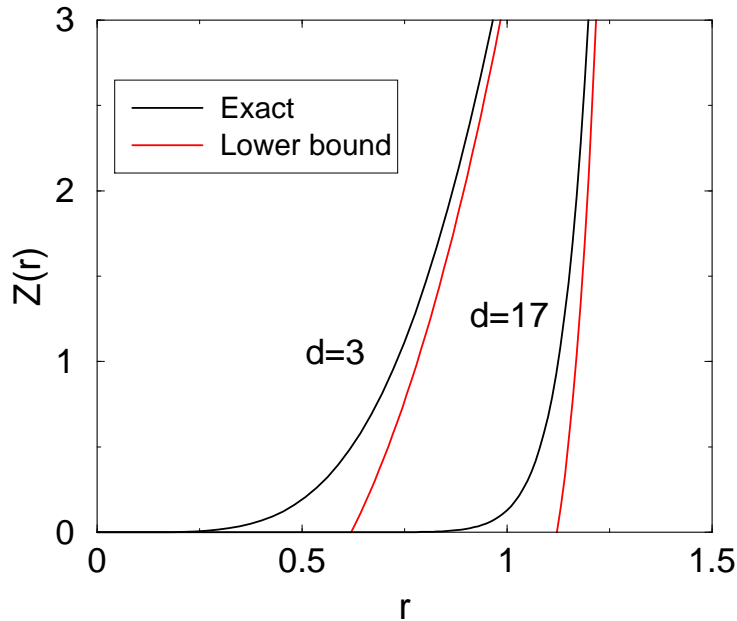,width=3.8in}}
\caption{Cumulative coordination number $Z(r)$ as a function of $r$
for $d=3$ and $d=17$ compared to the corresponding lower bounds
obtained from the inequality (\ref{Z-bounds}).}
\label{Z-plot}
\end{figure}

The results immediately above enable us 
to obtain the following small-$r$ and large-$r$ forms of the
cumulative coordination number:
\begin{eqnarray}
\fl Z(r)= v_1(r)\left[\frac{dK^2}{(d+2)^2} r^{2} - \frac{d(d+3)K^4}{2(d+2)^2(d+4)^2} r^{4} +\; {\cal O}(r^6)\right]\qquad (r \rightarrow 0)
\label{Z-small}
\end{eqnarray}
and
\begin{eqnarray}
Z(r)= v_1(r) - 1 + {\cal O}(r^{-1})  \qquad (r \rightarrow \infty).
\label{Z-large}
\end{eqnarray}
It is straightforward to prove that the asymptotic result (\ref{Z-large})
is a rigorous lower bound on $Z(r)$ for all $r$ using the 
identity $\int_0^\infty dx J_{d/2}^2(xK)/x=1/d$, the inequality
$\int_0^\infty dx J_{d/2}^2(xK)/x \ge \int_0^r dx J_{d/2}^2(xK)/x$ for
all positive $r$, and relation (\ref{Z-2}). Indeed, since $Z(r)$ is always nonnegative, we have
the lower bounds
\begin{eqnarray}
%Z(r) \ge \Bigg\{{0, \qquad\qquad r<D,\atop{v_1(r)-1, \quad r \ge D,}}
Z(r) \ge \left\{
\begin{array}{lr}
0, \quad r<D,\\
v_1(r)-1, \quad r\ge D,
\end{array}\right.
\label{Z-bounds}
\end{eqnarray}
where the length scale 
\begin{eqnarray}
D= \frac{\Gamma(1+d/2)^{1/d}}{\sqrt{\pi}}
\label{fermihole}
\end{eqnarray}
is the zero of $v_1(r)-1$. The length scale $D$ can be regarded
as an upper bound on the effective {\it hard-core diameter}, which
clearly grows with increasing dimension as vividly illustrated
in Fig. \ref{Z-plot}, which compares the exact form of $Z(r)$
with the lower bound (\ref{Z-bounds}) for $d=3$ and $d=17$.
For large $d$, the effective hard-core diameter is given
by the asymptotic expression
\begin{eqnarray}
D = \sqrt{\frac{d}{{2\pi e }}} \left[1+ \frac{\ln(d)}{2} +{\cal O}(1) \right] \qquad (d \rightarrow \infty),
\label{D-asym}
\end{eqnarray}
which is seen to grow like the square root of $d$. This growth
of the effective hard-core diameter with dimension is a conceptually
important conclusion that runs counter to conventional understanding
of corresponding fermionic systems, which identifies the inverse ``Fermi"
radius, i.e., $K^{-1}=1/[2\sqrt{\pi}\Gamma(1+1/d)^{1/d}]$
(a decreasing function of $d$) with an effective hard-core diameter.
We elaborate on this point in Section 3.

The hyperuniformity of the point process implies that the number variance
$\sigma^2(R)$ for a spherical window of radius $R$ must grow slower
than $R^d$ (i.e., the window volume) for large $R$. However, the fact that
the large-$r$ behavior of the pair correlation function is controlled
by the power law  $1/r^{d+1}$ means that $\sigma^2(R)$  must also grow faster than the surface area
of the window or $R^{d-1}$. In particular, upon substitution of (\ref{g2-d-2}) into (\ref{variance}),
an asymptotic analysis reveals that for large $R$
\begin{eqnarray}
\fl \sigma^2(R)= \Big\{\frac{d\pi^{(d-4)/2}}{2\Gamma[(d+1)/2]\Gamma(1+d/2)^{1/d}} \ln(R)+ C(d) \Big\}R^{d-1}\nonumber\\ 
+ {\cal O}(R^{d-2}) \qquad (R \rightarrow \infty),
\label{fermion-var}
\end{eqnarray}
where $C(d)$ is a $d$-dimensional constant of order unity. 
We remark that a similar asymptotic scaling is expected to hold even when the observation window is non-spherical; 
a discussion of this point has been provided in \cite{GiKl06}.
We see that the number variance
scaled by the window surface area, $\sigma^2(R)/R^{d-1}$, grows like $\ln(R)$, independent of the dimension. 
This unusual number variance growth law in three dimensions has also been
recently seen in  maximally random jammed sphere packings \cite{Do05}, which can be viewed
as  prototypical glasses because they are simultaneously perfectly rigid mechanically
and maximally disordered. Note that the coefficient multiplying
$\ln(R)$ in  (\ref{fermion-var}) decays to zero exponentially fast as $d \rightarrow \infty$,
and, therefore, the surface-area term $R^{d-1}$ increasingly
becomes the dominant one in the large-$d$ limit. 
This behavior should be contrasted with number
variance for a Poisson point process, which grows like the window volume, i.e., $R^{d}$.

\begin{figure}[bthp]
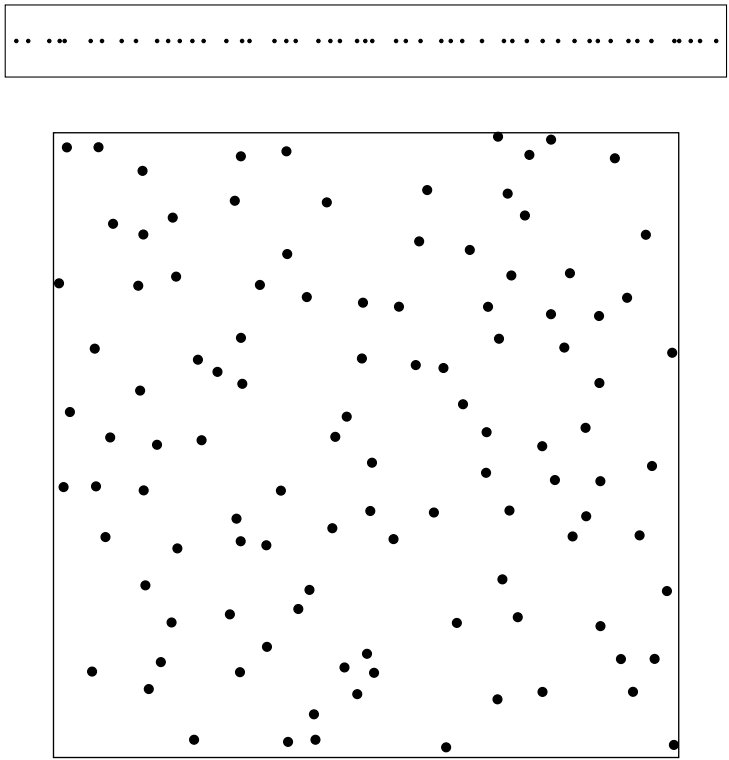

\centerline{\psfig{file=1D-50part.eps,width=3.8in,clip=}}\vspace{0.25in}
\centerline{\psfig{file=2D-100part.eps,width=3.3in,clip=}}
\caption{Top panel: A realization of 50 points of a Fermi-sphere point
process in a linear ``box"
subject to periodic boundary conditions. 
Bottom panel: A realization of 100 points of a Fermi-sphere point
process in a square box subject to periodic boundary conditions.}
\label{points}
\end{figure}

Figure \ref{points} graphically depicts the Fermi-sphere point processes 
in one and two dimensions, which are generated using the algorithm of Hough et. al. \cite{Ho06}.
Details and applications of this algorithm are reported
by us elsewhere \cite{Sc08}.

\subsection{``Fermi-Shells"  Point Processes in $\mathbb{R}^d$}

Here we consider a generalization of the Fermi-sphere point process in which $\tilde{H}(k)$ is an indicator
function for concentric rings in reciprocal space; we denote the resulting determinantal process as
the ``Fermi-shells'' point process.
Without loss of generality, define $2 m$ radii $k_F^{(j)}$, where $j \in \{1, 2, \ldots, 2m-1, 2m\}$ and $m \in \mathbb{N}$, 
such that the region of $d$-dimensional space within the ball $B(0; k_F^{(1)})$ is empty, the annulus between $k_F^{(2)}$ and $k_F^{(1)}$ is filled, and so forth.  
We therefore have $m$ filled concentric
rings in reciprocal space.  Since the conditions for a determinantal point process
are fulfilled for any indicator function in reciprocal space \cite{Cos04},
the pair correlation function will still be given by \eref{g2-determ}.  
The calculation of $H(r)$ proceeds as follows:
\begin{eqnarray}
H(r) = \left(\frac{1}{2\pi}\right)^d \sum_{j = 1}^{m}\left[\mathfrak{F}\{\Theta(k_F^{(2j)}-k)\}-\mathfrak{F}\{\Theta(k_F^{(2j-1)}-k)\}\right]\label{apone},
\end{eqnarray}
which implies:
\begin{eqnarray}
\fl H(r) = \left(\frac{1}{2\pi}\right)^{d/2}\sum_{j=1}^{m}\left[\left(\frac{k_F^{(2j)}}{r}\right)^{d/2} J_{d/2}(k_F^{(2j)}r)
-\left(\frac{k_F^{(2j-1)}}{r}\right)^{d/2} J_{d/2}(k_F^{(2j-1)} r)\right]\label{aptwo},
\end{eqnarray}
where $\mathfrak{F}$ denotes the Fourier transform to coordinate space.

\begin{figure}[bthp]
\centerline{\psfig{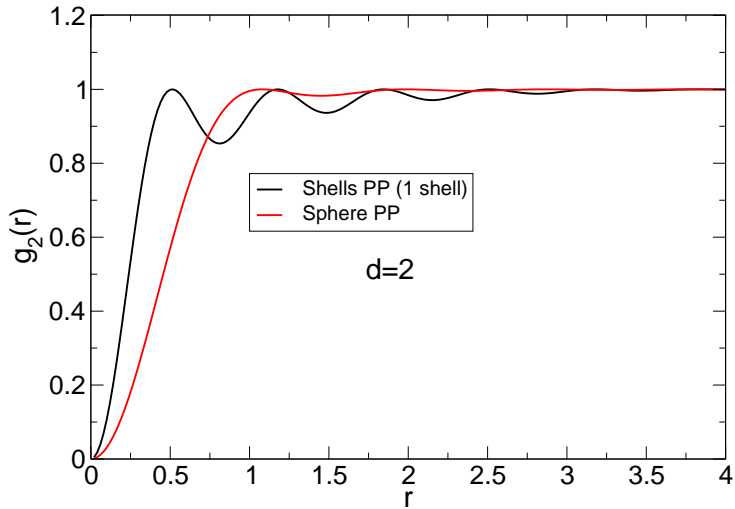}}
\caption{Comparison of the pair correlation function of the Fermi-sphere point process (PP) to that of the Fermi-shells
point process for $d=2$. In the latter case, $k_F=2\sqrt{\pi}$, $k_{F}^{(1)}=4$ and
$k_{F}^{(2)}=\sqrt{k_F^2+(k_{F}^{(1)})^2}$.
}
\label{donut}
\end{figure}

It is important to note that the values of the various $k_F^{(j)}$ are 
not independent of each other and are constrained by the density 
(here set to unity) according to:
\begin{eqnarray}
\sum_{j=1}^{m}\left[\left(k_F^{(2j)}\right)^d-\left(k_F^{(2j-1)}\right)^d\right] = (2\sqrt{\pi})^d \Gamma(1+d/2).
\end{eqnarray}
The filling of Fermi shells generally introduces a greater level
of short-range correlations relative to the Fermi-sphere case
(see Fig. \ref{donut}).

\subsection{Spin-Polarized Fermionic Gas in $\mathbb{R}^d$}

Note that we do not know of any correspondence of the general
Fermi-shells point
process in $\mathbb{R}^d$ for any $d \ge 2$ to random matrix theory
or the zeros of any generalized zeta function in number
theory. However, we can show that a spin-polarized
fermionic gas in $\mathbb{R}^d$ for any $d\ge 1$
has the same $n$-particle densities. For simplicity,
this comparison will be done for the Fermi-sphere case, i.e., we
will demonstrate that the $\rho_n$ are given by
(\ref{rhon-det}) with $H(r)$ given by (\ref{H2})
provided that the Fermi radius $k_F=K=2\sqrt{\pi}\Gamma(1+d/2)^{1/d}$.

We first recall some general properties of the $n$-particle density functions $\rho_n$ for a ground-state noninteracting gas of fermions in $\mathbb{R}^d, d\geq 1$.  Assume that we have $N$ spin-polarized fermions in a volume $V$ with number density $\rho = N/V$.  The $n$-particle density functions are defined for all $n \leq N$ with respect to the ground state $\vert\phi_0\rangle$ by:
\begin{eqnarray}\label{one}
\rho_n(\mathbf{x}_1, \ldots, \mathbf{x}_n) = \langle\phi_0| \prod_{i=1}^{n}\left[\psi^*(\mathbf{x}_i)\psi(\mathbf{x}_i)\right]|\phi_0\rangle,
\end{eqnarray}
where the operators $\psi^*(\mathbf{x}), \psi(\mathbf{x})$ are the creation and annihilation field operators, respectively.  Using the momentum representation, the field operators are defined in terms of the particle creation and annihilation operators $a_{\mathbf{k}}^*, a_{\mathbf{k}}$:
\begin{eqnarray}
\psi(\mathbf{x}) = \sum_{\mathbf{k}} \phi_{\mathbf{k}}(\mathbf{x}) a_{\mathbf{k}}\\
\psi^*(\mathbf{x}) = \sum_{\mathbf{k}} \overline{\phi_{\mathbf{k}}(\mathbf{x})} a^*_{\mathbf{k}}\\
\phi_{\mathbf{k}}(\mathbf{x}) = \left(\frac{1}{V}\right)^{1/2} \exp\left[\rmi(\mathbf{k},\mathbf{x})\right]\\
\overline{\phi_{\mathbf{k}}(\mathbf{x})} = \left(\frac{1}{V}\right)^{1/2} \exp\left[-\rmi(\mathbf{k}, \mathbf{x})\right],
\end{eqnarray}
where $(\mathbf{k}, \mathbf{x}) = \sum_{i=1}^{d} k_i x_i$ is the Euclidean inner product of two real-valued vectors.
The field operators for fermions must satisfy the following anticommutation relation:
\begin{eqnarray}\label{two}
\left\{\psi(\mathbf{x}_i), \psi^*(\mathbf{x}_j)\right\} = \delta(\mathbf{x}_i - \mathbf{x}_j).
\end{eqnarray}

Wick's Theorem along with the anticommutation relation \eref{two} immediately allow us to write the expectation value in \eref{one} as a determinant (equivalently, see Macchi's discussion of fermion processes \cite{Ma75}):
\begin{eqnarray}
\rho_n(\mathbf{x}_1, \ldots, \mathbf{x}_n) &= \sum_{\sigma \in S_n} (-1)^{\sigma} \prod_{i=1}^{n} \langle\phi_0| \psi^*(\mathbf{x}_i)\psi(\mathbf{x}_{\sigma(i)})|\phi_0\rangle\label{seven}\\
&= \det\left[\langle\phi_0| \psi^*(\mathbf{x}_i)\psi(\mathbf{x}_{j})|\phi_0 \rangle\right]_{1\leq i, j \leq n}\label{eight},
\end{eqnarray}
where $S_n$ denotes the permutation group for $n$ objects.  Namely, Wick's Theorem relates the $n$-particle density functions to the $n!$ ways of pairing the $2n$ creation and annihilation field operators; anticommutation of the operators introduces the factor of $(-1)^{\sigma}$ in \eref{seven}.  

For the ground-state system, the argument of the determinant in \eref{eight} may be evaluated by filling the Fermi sphere up to the Fermi radius $k_F$ with the result:
\begin{eqnarray}\label{nine}
G(\mathbf{x}, \mathbf{x}^{\prime}) = \langle\phi_0|\psi^*(\mathbf{x})\psi(\mathbf{x}^{\prime})|\phi_0\rangle = \left(\frac{k_F}{2\pi}\right)^{d/2} r^{-d/2} J_{d/2}(k_F|\mathbf{x}-\mathbf{x}^{\prime}|).
\end{eqnarray}
Since $G$ is a translationally-invariant function, we will abuse notation slightly and write $G(\mathbf{x}, \mathbf{x}^{\prime}) = G(|\mathbf{x}-\mathbf{x}^{\prime}|) = G(r; k_F),$ making the parameterization by $k_F$ explicit. 

In accordance with our established convention and without loss of generality, let $\rho = 1$.  For a system of $N$ spin-polarized fermions, the Fermi radius $k_F$ is given exactly by:
\begin{eqnarray}
k_F = 2\sqrt{\pi}\left\{\Gamma\left[(d/2)+1\right]\right\}^{1/d} = K.
\end{eqnarray}
Therefore, we may equivalently write for \eref{nine}:
%It is sufficient then to show $H(r) = G(r; K)$, where $K = 2\sqrt{\pi}\left\{\Gamma\left[(d/2)+1\right]\right\}^{1/d},$  to make the connection between noninteracting fermions and our proposed determinantal process.  Evaluating \eref{nine} for $k_F = K$ immediately yields the desired result:
\begin{eqnarray}\label{eleven}
G(r; K) =  \pi^{-d/4} \sqrt{\Gamma\left[(d/2)+1\right]} r^{-d/2} J_{d/2}(rK) = H(r).
\end{eqnarray}
The result in \eref{eleven} shows that our proposed Fermi-sphere point process corresponds \emph{exactly} 
to the one generated by a system of noninteracting fermions in $d$ dimensions.  

The connection for $d = 1$ between this system of noninteracting fermions and the CUE of random matrix theory implies that the correlations
in the ground state resulting from the Pauli exclusion principle are equivalent to those induced on a 
$d = 2$ Coulomb gas constrained to the unit ring
and interacting via a logarithmic pair potential at finite temperature $T = 1/2$ in units such that Boltzmann's constant $k_B = 1$.  The argument in Appendix A shows
that this reduction of the probability density to a classical particle system
with at most two-body interactions is peculiar to the choice of
the indicator function in reciprocal space; in general, one must include at least three-body interactions to describe the system appropriately.

We also mention the significance of \eref{fermihole} in defining an effective hard core on the system of noninteracting fermions
in each dimension.  This issue is equivalent to assigning an appropriate length scale for the
widely-studied (see, e.g., \cite{Mc60}) Fermi correlation hole.  
An argument primarily due to Slater \cite{Sl51} suggests that the correlations which result from antisymmetry in the many-body wavefunction 
extend outward for a distance $r_0 \lessapprox k_F^{-1}$.  The reasoning behind this choice
of length scale relies on the introduction of an exchange hole into the system with approximate spherical symmetry and a radius determined by
the de Broglie wavelength associated with the Fermi radius $k_F$.  This notion has been quantified for interacting atomic systems 
by considering, for example, the difference of distribution functions derived from Hartree-Fock and Hartree wavefunctions \cite{BoCo74},
but the $k_F^{-1}$ scaling for the noninteracting case is still reported in modern texts \cite{Sc05}.  

However, we recall from \eref{R} that as $d$ increases, the value of the Fermi radius also increases; in fact, $k_F \sim \sqrt{d}$ for sufficiently
large $d$.  This behavior implies that $k_F^{-1} \rightarrow 0$ as $d \rightarrow +\infty$, meaning that the 
effective hard core diameter would become negligibly small for sufficiently large $d$ if it were to scale as $k_F^{-1}$.  This conclusion 
contradicts the fact that $g_2(r) \rightarrow 0$ pointwise as $d \rightarrow +\infty$ for all $r \in [0, +\infty)$. In other words, the range in $r$ over which one finds a small probability of finding two particles 
in close proximity increases with dimension, requiring a different means by which to quantify the extent of the effective hard core.

The definition of $Z(r)$ in \eref{Z} suggests instead that we take the value of $D$ in \eref{fermihole} as a measure of the effective hard 
core diameter.  We note that since $Z(r)$ is a nonnegative monotonically increasing function of $r$ and, therefore, nonzero for some range in $r$ 
over which its lower bound is zero, $D$ must represent an upper bound to the effective hard core radius.  This representation of the effective
hard-core diameter is quantitatively well-defined for any dimension due to the inclusion of the 
$x^{d-1}$ factor under the integral in \eref{Z} from the
surface area of the $d$-dimensional ball; it is this factor which appropriately rescales $g_2$ such that \eref{Z-bounds} represents a true lower bound
on $Z(r)$ in any dimension by the argument above.  
It is also for this reason that neither $g_2(r)$ nor $S(k)$ alone are sufficient to define the effective hard-core diameter
in a quantitative manner.  In contrast to Slater's scaling of the effective hard core diameter, 
$D \sim k_F$ in any dimension by the definition of the Fermi radius, 
which is in accordance with the high-dimensional behavior of both $g_2$ and $k_F$.

\section{Nearest-Neighbor Functions}

It is useful to characterize point processes by examining other
quantities besides the $n$-particle correlation functions.
One popular descriptor used in one dimension is the so-called
``gap" distribution function $p(z)$ \cite{Me67}, which characterizes
the spacing between the points. In the random matrix 
theory literature, this quantity often has erroneously
and misleadingly been called the ``nearest-neighbor-spacing"
distribution because gaps to the right of some reference
point are considered. However, $p(z)$ makes no distinction
between gaps to the left or right of some reference point.
The quantity $p(z) dz$ gives the probability of finding
a gap (a line interval empty of points) of length  between
$z$ and $z + dz$. The function $p(z)$ is called the {\it chord length probability density}
in the theory of random media \cite{To93,To02a}.

In the case of random matrix theory, there exist exact
representations of $p(z)$ for the spacings of the eigenvalues
in the GOE, GUE, and GSE, but they can only
be determined numerically for general situations. A remarkably
accurate approximation for the GOE in the infinitely-large
matrix limit is the so-called Wigner surmise.
The Wigner surmise has been generalized to any of 
the aforementioned ensembles as follows:
\begin{eqnarray}
p_{\beta}(z)=A_{\beta} z^{\beta} e^{-B_{\beta} z^2},
\label{wigner}
\end{eqnarray}
where the parameters $A_{\beta}$ and $B_{\beta}$, which
depend on reciprocal temperature $\beta$, are obtained
from the normalization of both $p(z)$ and its first moment, or the average gap size
$\langle z\rangle$.
For the GOE, GUE, and GSE, $\beta=1, 2$, and 4, respectively,
and $A_1=\pi/2$, $B_1=\pi/4$, $A_2=32/\pi^2$, $B_2=4/\pi$, and
$A_4=262144/(729\pi^2)$, $B_4=64/(9\pi)$.

For $d \ge 2$, the gap distribution function is strictly not a meaningful
descriptor of point processes. The natural generalizations of $p(z)$ in
higher dimensions are the nearest-neighbor functions \cite{To90,To02a}.
Nearest-neighbor functions describe the probability of finding
the {\it nearest point} of a point process in $\mathbb{R}^d$  at some given distance from a 
{\it reference point} in the space. Such statistical quantities are
called ``void" or ``particle" nearest-neighbor functions 
if the the reference point is an arbitrary
point of the space or an actual point of the point process, respectively.

\subsection{Definitions}

First, we recall the definitions of the void and particle nearest-neighbor probability
density functions $H_{V}(r)$ and $H_{P}(r)$, respectively:
\begin{eqnarray}
\begin{array}{ccp{3.7in}}
H_{V}(r)\,\rmd r & = & Probability that a point of the point process
lies at a distance between $r$ and $r + \rmd r$ from an arbitrary
point in the space. 
\end{array} 
\label{def-Hv}
\end{eqnarray}
\begin{eqnarray}
\begin{array}{ccp{3.7in}}
H_{P}(r)\,\rmd r & = & Probability that a point of the point process
lies at a distance between $r$ and $r + \rmd r$ from another 
point of the point process.
\end{array} 
\label{def-Hp}
\end{eqnarray}
It is useful to  introduce the associated dimensionless ``exclusion" probabilities $E_{V}(r)$ and  $E_{P}(r)$ defined as follows:
\begin{eqnarray}
\begin{array}{ccp{3.7in}}
E_{V}(r) & = & Probability of finding a spherical cavity
of radius $r$ empty of any points in the point process.
\end{array}
\label{def-Ev}
\end{eqnarray}
\begin{eqnarray}
\begin{array}{ccp{3.7in}}
E_{P}(r) & = & Probability of finding a spherical cavity
of radius $r$ centered at an arbitrary point of the point process
empty of any other points.
\end{array}
\label{def-Ep}
\end{eqnarray}

It follows that the exclusion probabilities are {\it complementary cumulative distribution
functions} associated with
 the density functions and thus are related
to the latter via
\begin{eqnarray}
E_{V}(r) = 1 - \int_{0}^{r} H_{V}(x) \, \rmd x
\label{Ev-cum}
\end{eqnarray}
and
\begin{eqnarray}
E_{P}(r) = 1 - \int_{0}^{r} H_{P}(x) \, \rmd x.
\label{Ep-cum}
\end{eqnarray}
Differentiating the exclusion-probability relations with
 respect to $r$ gives
\begin{eqnarray}
H_{V}(r) = -\frac{\partial E_{V}}{\partial r}
\label{derv-Ev}
\end{eqnarray}
and
\begin{eqnarray}
H_{P}(r) = -\frac{\partial E_{P}}{\partial r}.
\label{derv-Ep}
\end{eqnarray}
The $n$th moment of $H_P(r)$ is defined as
\begin{eqnarray}
\lambda_n= \int_{0}^{\infty} r^n H_P(r) \, \rmd r \label{nth-mom}.
\end{eqnarray} 
Of particular interest to us is the {\it mean nearest-neighbor distance}
\begin{eqnarray}
\lambda \equiv \lambda_1&= \int_{0}^{\infty} r H_P(r) \rmd r \nonumber \\
&= \int_{0}^{\infty} E_P(r) \rmd r.
\label{lambda}
\end{eqnarray}

It is useful to express the density functions $H_V(r)$ and $H_P(r)$ as
a product of two functions as follows:
\begin{eqnarray}
H_{V}(r) = \rho s_{1}(r) G_{V}(r) E_{V}(r),
\label{def-Hv2}
\end{eqnarray}
and
\begin{eqnarray}
H_{P}(r) = \rho s_{1}(r) G_{P}(r) E_{P}(r),
\label{def-Hp2}
\end{eqnarray}
where $s_1(r)$ is the surface area of a  
$d$-dimensional sphere of radius $r$ given by (\ref{area-sph}).
The quantities $G_V(r)$ and $G_P(r)$ are called the ``conditional" nearest-neighbor
functions and have the following interpretation:
\begin{eqnarray}
\begin{array}{ccp{3.1in}}
\rho s_{1}(r) G_{V}(r) \rmd r & = &
Given that a spherical cavity of radius $r$ centered
at an arbitrary point in the space is
empty of any points of the point process, the probability of finding
a point in the spherical shell of volume
$s_{1}(r)\rmd r$ surrounding the arbitrary point.
\end{array}
\label{def-Gv}
\end{eqnarray}
\begin{eqnarray}
\begin{array}{ccp{3.1in}}
\rho s_{1}(r) G_{P}(r) \rmd r & = & 
Given that a spherical cavity of radius $r$ centered 
at a randomly selected point of the point process is
empty of any other points, the probability of finding
a point  in the spherical shell of volume
$s_{1}(r)\rmd r$ surrounding the randomly selected point. 
\end{array}
\label{def-Gp}
\end{eqnarray}
Thus, it follows that the exclusion probabilities are also given by
\begin{eqnarray}
E_V(r)=\exp\left[-\rho s_1(1) \int_0^r x^{d-1} G_V(x) \rmd x\right]
\label{EV-GV}
\end{eqnarray}
and
\begin{eqnarray}
E_P(r)=\exp\left[-\rho s_1(1) \int_0^r x^{d-1} G_P(x) \rmd x\right].
\label{EP-GP}
\end{eqnarray}
We remark that knowledge of any one function $H$, $E$, or $G$ (either void or particle) is sufficient
to determine the other two functions via the relations mentioned above.

\subsection{Series Representations}

The nearest-neighbor functions can be expressed as infinite series
whose terms are integrals over the $n$-particle density functions \cite{To90,To02a}.
For example, the void and particle exclusion probability functions for a translationally
invariant point process are respectively given by
\begin{eqnarray}
E_{V}(r) = 1+ \sum^{\infty}_{k=1} (-1)^{k} \frac{\rho^k}{k!}
 \int_{{\mathbb R}^d} g_{k}({\bf r}^{\,k}) \prod_{j=1}^k \Theta(r-|{\bf x} - {\bf r}_{j}|) \rmd{\bf r}_j
\label{Ev-series}
\end{eqnarray}
and
\begin{eqnarray}
E_P (r) = 1+\sum_{k=1}^{\infty} (-1)^{k} \frac{\rho^k}{k!} \int_{{\mathbb R}^d} g_{k+1}(\mathbf{r}^{k+1}) \prod_{j=2}^{k+1} 
\Theta(r-r_{1j}) \rmd{\bf r}_j.
\label{Ep-series}
\end{eqnarray}
The corresponding series for $H_V(r)$ and $H_P(r)$ 
are obtained from the series above using (\ref{derv-Ev}) and (\ref{derv-Ep}).

In general, an exact evaluation of the aforementioned infinite series are not possible, except for 
simple processes such as the Poisson point process. In the latter instance,
because $\rho_n=\rho^n$, both series (\ref{Ev-series}) and (\ref{Ep-series}) can be summed
exactly to give
\begin{eqnarray}
E_V(r)=E_P(r)=\exp[-\rho v_1(r)],
\label{E-poisson}
\end{eqnarray}
where $v_1(r)$ is the volume of a $d$-dimensional sphere of radius $r$ given by (\ref{v1}).
Therefore, for a Poisson point process, we have from (\ref{derv-Ev}), (\ref{derv-Ep}), 
(\ref{def-Hv2}), and (\ref{def-Hp2}) that 
\begin{eqnarray}
H_V(r)=H_P(r)=\rho s_1(r) \exp[-\rho v_1(r)], \qquad G_V(r)=G_P(r)=1.
\end{eqnarray}
We see that there is no distinction between the void and particle quantities
for the Poisson distribution, which is generally not the case
for correlated point processes.

\subsection{Rigorous Bounds}

Torquato has given rigorous upper and lower bounds on
 the so-called  {\it canonical $n$-point correlation function} 
$H_{n} ({\bf x}^{m}; {\bf x}^{p-m}; {\bf r}^{q})$ (with $n=p+q$ and $m \le p$) for 
point processes in $\mathbb{R}^d$.  Since the void and  particle exclusion probabilities and nearest-neighbor probability density functions are just special cases of $H_{n}$,
 then we also have strict bounds on them for such models.
 Let $X$ represent either $E_{V}, H_{V}, E_{P}$, or $H_{P}$
 and $X^{(k)}$ represent the $k$th term of the series
for these functions. Furthermore, let
\begin{eqnarray}
W^{\ell}  =  \sum_{k=0}^{\ell} (-1)^{k} X^{(k)} \\
\end{eqnarray}
be the partial sum.  Then it follows   that
 for any of the exclusion probabilities or nearest-neighbor
 probability density functions, we have the bounds
\begin{eqnarray}
X & \leq & W^{\ell}, \qquad \mbox{for $\ell$ even}
\mbox{} \nonumber \\
X & \geq & W^{\ell}, \qquad \mbox{for $\ell$ odd}.
\end{eqnarray}

Application of the aforementioned inequalities yield the 
first three successive bounds on the nonnegative void exclusion probability:
\begin{eqnarray}
E_V(r) \le 1 \\
E_V(r) \ge 1 -\rho v_1(r) \label{EV-bound1}\\
E_V(r)  \le  1 -\rho v_1(r) +\frac{\rho^2}{2} s_1(1)\int_0^{2r} x^{d-1} v_2^{int}(x;r) g_2(x) \rmd x
\label{EV-bound2},
\end{eqnarray}
where $v_2^{int}(x; r) = v_1(r)\alpha(x;r)$ is the intersection volume of two $d$-dimensional spheres (cf. \eref{alpha1}).
The corresponding first two nontrivial bounds on the 
nonnegative nearest-neighbor probability density function $H_V(r)$
are as follows:
\begin{eqnarray}
H_V(r) \le \rho s_1(r) \label{HV-bound1}\\
H_V(r)  \ge  \rho s_1(r) -
\frac{\rho^2}{2} s_1(1)\int_0^{2r} x^{d-1} s_2^{int}(x;r) g_2(x) \rmd x,
\label{HV-bound2}
\end{eqnarray}
where $s_2^{int}(x;r)\equiv \partial v_2^{int}(x;r)/\partial r$
is the surface area of the intersection volume $v_2^{int}(x;r)$.
Bounds on the conditional 
function $G_V(r)$ follow by combining the bounds above on $E_V(r)$ and $H_V(r)$ and definition
(\ref{def-Hv2}). For example, we obtain the following bounds
\begin{eqnarray}
G_V(r) \le \frac{1}{1-\rho v_1(r)}
\label{GV-bound1}
\end{eqnarray}
and 
\begin{eqnarray}
G_V(r) \ge \frac{1 - \frac{\rho}{s_1(r)} s_1(1)\int_0^{2r} x^{d-1} s_2^{int}(x;r) g_2(x) \rmd x}
{1 -\rho v_1(r) +\frac{\rho^2}{2} s_1(1)\int_0^{2r} x^{d-1} v_2^{int}(x;r) g_2(x) \rmd x},
\label{GV-bound2}
\end{eqnarray}
which should only be applied for $r$ such that $G_V(r)$ remains positive.
The bounds above lead to the following properties of the nearest-neighbor
functions at the origin:
\begin{eqnarray}
E_V(0)=1, \qquad H_V(0)=0, \qquad G_V(0)=1.
\end{eqnarray}

Similarly, the first three successive bounds on the particle exclusion probability
are given by
\begin{eqnarray}
E_P(r) \le 1 \\
E_P(r) \ge 1 - Z(r) \label{EP-bound1}\\
E_P(r) \le 1 - Z(r)\nonumber\\
+ \frac{\rho^2}{2} \int_{{\mathbb R}^d}  \int_{{\mathbb R}^d} \Theta(r- r_{12})\Theta(r - r_{13}) g_3(r_{12},r_{13},r_{23}) \rmd{\bf r}_2 \rmd{\bf r}_3,
\label{EP-bound2}
\end{eqnarray}
where $Z(r)$ is the cumulative coordination number defined by (\ref{Z}).
The corresponding first two nontrivial bounds on the 
nonnegative nearest-neighbor probability density function $H_P(r)$
are as follows:
\begin{eqnarray}
H_P(r) \le \rho s_1(r)g_2(r) \label{HP-bound1}\\
H_P(r) \ge  \rho s_1(r) g_2(r)\nonumber\\ 
-\rho^2\int_{{\mathbb R}^d}  \int_{{\mathbb R}^d} \delta(r- r_{12})\Theta(r - r_{13}) g_3(r_{12},r_{13},r_{23}) \rmd{\bf r}_2 \rmd{\bf r}_3,
\label{HP-bound2}
\end{eqnarray}
where $\delta(r)$ is the radial Dirac delta function. Bounds on the conditional 
function $G_P(r)$ follow by combining the bounds on $E_P(r)$ and $H_P(r)$ and definition
(\ref{def-Hp2}). For example, we obtain the following bounds:
\begin{eqnarray}
G_P(r) \le \frac{g_2(r)}{1-Z(r)}
\label{GP-bound1}
\end{eqnarray}
and 
\begin{eqnarray}
\fl G_P(r) \ge \frac{g_2(r) - \frac{\rho}{s_1(r)} \int_{{\mathbb R}^d}  \int_{{\mathbb R}^d} \delta(r- r_{12})\Theta(r - r_{13}) g_3(r_{12},r_{13},r_{23}) \rmd{\bf r}_2 \rmd{\bf r}_3}{ 1 - Z(r)+ \frac{\rho^2}{2} \int_{{\mathbb R}^d}  \int_{{\mathbb R}^d} \Theta(r- r_{12})\Theta(r - r_{13}) g_3(r_{12},r_{13},r_{23}) \rmd{\bf r}_2 \rmd{\bf r}_3},
\label{GP-bound2}
\end{eqnarray}
which should only be applied for $r$ such that $G_P(r)$ remains positive.
The bounds above lead to the following properties of the nearest-neighbor
functions at the origin:
\begin{eqnarray}
E_P(0)=1, \qquad H_P(0)=0, \qquad G_P(0)=0.
\end{eqnarray}

We now obtain bounds on the mean nearest-neighbor distance
$\lambda$ at unit density using the aforementioned upper
and lower bounds on $E_P(r)$. Let us define the following
distances:
\begin{eqnarray}
\lambda_L= \int_0^{r_0} [1- Z(r)] \rmd r
\label{L1}
\end{eqnarray}
and
\begin{eqnarray}
\lambda_U= \int_0^{+\infty} \exp[-Z(r)] \rmd r,
\label{L2}
\end{eqnarray}
where $r_0$ is the location of the zero of $1- Z(r)$.
In light of the bounds (\ref{EP-bound1}) and (\ref{EP-GP-bound}), it is clear
that $\lambda_L$ and $\lambda_U$ bound $\lambda$ from
below and above, respectively, i.e.,
\begin{eqnarray}
\lambda_L \le \lambda \le \lambda_U.
\end{eqnarray}

\subsection{Results for Fermi-Sphere Point Processes}

\subsubsection{Exact Determinantal Representations}

The Fermi-sphere point process is unique in that both the 
particle and void exclusion probabilities may be expressed as determinants over $N\times N$ matrices,
the elements of which are related to overlap integrals of the basis functions 
$\phi_{\mathbf{k}} = (1/\sqrt{V})\exp[\rmi (\mathbf{k},\mathbf{x})]$
on $B(\mathbf{0}; r)$,  
a $d$-dimensional ball of radius $r$ centered at the origin (the exact location of the ball's center
is irrelevant since the point process is translationally invariant).  
The thermodynamic limit can then be taken appropriately.    
We provide the details of this analysis elsewhere \cite{Sc08} and immediately state the results:
\begin{eqnarray}
E_V(r) = \det[\mathbb{I}-M(r)]\label{det1}\\
E_P(r) = E_V(r)\tr\{A[\mathbb{I}-M(r)]^{-1}\}\label{det2},
\end{eqnarray}
where $\mathbb{I}$ is the $N\times N$ identity matrix, and the matrices $M(r)$ and $A$ are defined by:
\begin{eqnarray}
M_{ij}(r) = \int_{B(\mathbf{0};r)} \overline{\phi_i(\mathbf{x})}\phi_j(\mathbf{x}) \rmd\mathbf{x}\label{det3}\\
A_{ij} = \overline{\phi_i(\mathbf{0})}\phi_j(\mathbf{0})/\rho\label{det4},
\end{eqnarray}
where $\rho$ is the number density.
We recall that knowledge of $E_V(r)$ and $E_P(r)$ is sufficient to
determine all of the remaining nearest-neighbor functions $H_{V/P}(r)$ and $G_{V/P}(r)$.
Note that as $r\rightarrow 0$, $M_{ij}(r) \rightarrow 0$ for all $i$ and $j$, which provides the
necessary results $E_V(0) = 1$ and $E_P(0) = \tr(A) = \sum_i|\phi_i(\mathbf{0})|^2/\rho = H(\mathbf{0}) = 1.$ We mention that these determinants must be evaluated numerically but essentially
exactly for finite $N\times N$ matrices, where $N$ is chosen to be
sufficiently large to capture accurately the behavior of the system in the thermodynamic limit.  Evidence for this
convergence is provided in another paper \cite{Sc08}.

\subsubsection{Bounds, Comparison to Exact Results, and Link to Sphere Packings}

We now obtain bounds for nearest-neighbor functions for the Fermi-sphere point process
and compare them to the corresponding aforementioned exact results. 
%Unless otherwise stated, we will set the number density $\rho$ to be unity.
Using the identity 
\begin{eqnarray}
s_1(1)\int_0^{2r} x^{d+1} v_2^{int}(x;r) \rmd x= \frac{2d}{d+2} r^2 [v_1(r)]^2,
\end{eqnarray}
the leading order term of the small-$r$ expansion (\ref{quad}),
and the upper bound (\ref{EV-bound2}), we obtain the weaker upper bound
\begin{eqnarray}
E_V(r) \le 1 - \rho v_1(r) +  \rho^2\frac{dK^2}{(d+2)^2} r^2 [v_1(r)]^2,
\end{eqnarray}
which is exact through terms of  order $r^{2(d+1)}$.
Therefore, we also have
\begin{eqnarray}
H_V(r) = \rho s_1(r)  - \rho^2\frac{2(d+1)  K^2}{d(d+2)^2}  r^3 [s_1(r)]^2  +
{\cal O}(r^{2d+3}) 
\end{eqnarray}
and 
\begin{eqnarray}
G_V(r) = 1 +\rho v_1(r) +\rho^2 v_1(r)^2  +{\cal O}(r^{2d+1}) .
\end{eqnarray}
We see that through order $r^{2d+1}$, $G_V(r) \ge 1$. This bound
will be shown to apply for any $r$.

Employing the inequality $ g_3(r_{12},r_{13},r_{23})\le g_2(r_{12})g_2(r_{13})$ 
[cf. (\ref{rhon-ineq})] in the upper bound (\ref{EP-bound2}) on $E_V(r)$ and lower bound
(\ref{HP-bound2}) on $H_P(r)$, we find the following
corresponding weaker bounds:
\begin{eqnarray}
E_P(r) \le 1 - Z(r)+ \frac{Z^2(r)}{2}
\end{eqnarray}
and 
\begin{eqnarray}
H_P(r) \ge \rho s_1(r) g_2(r) - \rho s_1(r) g_2(r) Z(r).
\end{eqnarray}
These bounds in conjunction with the analogous evaluations
of the bounds (\ref{GP-bound1}) and (\ref{GP-bound2}) on the conditional
pair function $G_P(r)$ yields its exact small-$r$
behavior up through terms of order $r^4$:
\begin{eqnarray}
G_P(r)= \frac{K^2}{d+2} r^2 - \frac{(d+3)K^4}{2(d+2)^2(d+4)} r^4 +\; {\cal O}(r^6).
\end{eqnarray}
Comparing this expansion to (\ref{quad}) reveals that $G_P(r)=g_2(r)$
up through terms of order $r^4$. 

We can also show that 
\begin{eqnarray}\label{GvLL}
G_V(r) \ge 1 \qquad \mbox{for all} \quad r
\end{eqnarray}
and
\begin{eqnarray}\label{GpLL}
G_P(r) \ge g_2(r) \qquad \mbox{for all} \quad r.
\end{eqnarray}
These results are obtained from definitions (\ref{EV-GV}) and (\ref{EP-GP}) and
the following upper bounds on the exclusion probabilities:
\begin{eqnarray}
E_V(r)\leq \exp[-\rho v_1(r)] \qquad \mbox{for all} \quad r
\label{EV-GV-bound}
\end{eqnarray}
and
\begin{eqnarray}
E_P(r)\leq \exp[- Z(r)] \qquad \mbox{for all} \quad r.
\label{EP-GP-bound}
\end{eqnarray}
To prove \eref{EV-GV-bound} and \eref{EP-GP-bound}, we recall the series 
representations of $E_V$ and $E_P$ in \eref{Ev-series} and \eref{Ep-series}, respectively,
which we may rewrite in the following more compact form:
\begin{eqnarray}
E_{P/V}(r) = 1+\sum_{k=1}^{+\infty} (-1)^k E_{P/V}^{(k)}\label{Epvcompact}\\
E_V^{(k)} = \frac{\rho^k}{k!} \int_{\mathbb{R}^d} g_k(\mathbf{r}^k) \prod_{j=1}^k \Theta(r-|\mathbf{x}-\mathbf{r}_j|) \rmd\mathbf{r}_j\label{Evk}\\
E_P^{(k)} = \frac{\rho^k}{k!} \int_{\mathbb{R}^d} g_{k+1}(\mathbf{r}^k) \prod_{j=2}^{k+1}\Theta(r-r_{1j})\rmd\mathbf{r}_j\label{Epk}.
\end{eqnarray}
It is important to note that the series in \eref{Epvcompact} converge absolutely
for all $r \in \mathbb{R}^+$, which is easily seen from the 
inequalities
\begin{eqnarray}
E_V^{(k)} \leq \frac{\rho^k}{k!} \int_{\mathbb{R}^d} \prod_{j=1}^{k} \Theta(r-|\mathbf{x}-\mathbf{r}_j|)\rmd\mathbf{r}_j
= \frac{[\rho v_1(r)]^k}{k!}\label{Evkineq}\\
E_P^{(k)} \leq \frac{\rho^k}{k!} \int_{\mathbb{R}^d} \prod_{j=2}^{k+1} g_2(r_{1j})\Theta(r-r_{1j}) \rmd\mathbf{r}_j
= \frac{[Z(r)]^k}{k!} \label{Epkineq}.
\end{eqnarray}
Equations \eref{Evkineq} and \eref{Epkineq} follow directly from the inequalities in \eref{rhotriv} and \eref{rhon-ineq}, respectively. 
It is therefore true that
\begin{eqnarray}
| E_V(r)| \leq 1+\sum_{k=1}^{+\infty}E_V^{(k)} \leq \sum_{k=0}^{+\infty}\frac{[\rho v_1(r)]^k}{k!} = \exp[\rho v_1(r)] < +\infty\\
| E_P(r)| \leq 1+\sum_{k=1}^{+\infty}E_P^{(k)} \leq \sum_{k=0}^{+\infty}\frac{[Z(r)]^k}{k!} = \exp[Z(r)] < +\infty,
\end{eqnarray}
and absolute convergence of the series in \eref{Epvcompact} implies convergence of those series, which in turn allows us to conclude
that the sequences $E_{P/V}^{(k)}\rightarrow 0$ as $k \rightarrow +\infty$.  

We now wish to compare the series in \eref{Epvcompact} with the following series representations of the proposed upper limits in \eref{EV-GV-bound} and 
\eref{EP-GP-bound}:
\begin{eqnarray}
E_V^{(UL)}(r) = 1-\rho v_1(r) + \sum_{k=2}^{+\infty} \frac{[-\rho v_1(r)]^k}{k!}\label{EvULseries}\\
E_P^{(UL)}(r) = 1-Z(r)+\sum_{k=2}^{+\infty} \frac{[-Z(r)]^k}{k!}\label{EpULseries}.
\end{eqnarray}
Note that the series in \eref{Epvcompact}, \eref{EvULseries}, and \eref{EpULseries} agree up to their second terms; the alternating series test
then implies that the contributions from the remaining terms is no greater than the magnitude of the third terms in the series.
Equations \eref{Evkineq} and \eref{Epkineq} therefore show that the series in \eref{EvULseries} and \eref{EpULseries} in actuality do bound the
series in \eref{Epvcompact} from above, thereby proving the claims.  
%We stress that convergence of the series \eref{Epvcompact} is crucial
%for these bounds to hold.
The lower bounds in \eref{GvLL} and \eref{GpLL} immediately follow from monotonicity and positivity of the exponential function along with
\eref{EV-GV} and \eref{EP-GP}.

\begin{figure}[H]
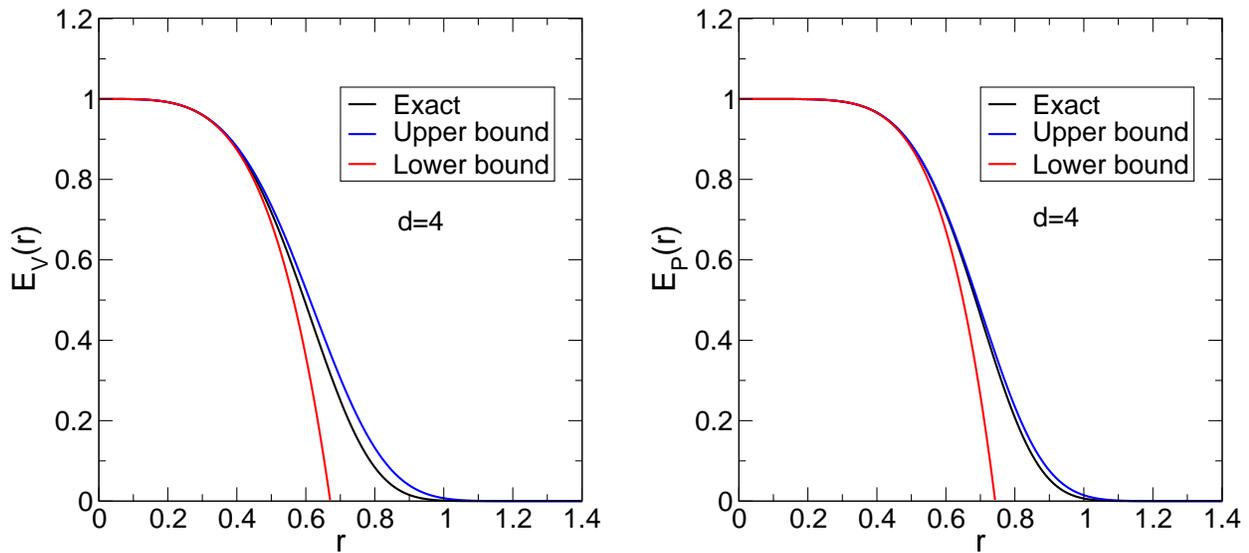

\centerline{\psfig{file=EV-4D-Bounds.eps,width=3.1in,clip=}\hspace{0.25in}\psfig{file=EP-4D-Bounds.eps,width=3.1in,clip=}}
\caption{Left panel: Upper and lower bounds on $E_V(r)$ for $d=4$ and $\rho=1$
as obtained from (\ref{EV-bound1}) and (\ref{EV-GV-bound}) 
compared to the corresponding exact evaluation of it. 
Right panel: Upper and lower bounds on $E_P(r)$ for $d=4$ and $\rho=1$
as obtained from (\ref{EP-bound1}) and (\ref{EP-GP-bound}), respectively, compared to the corresponding exact evaluation of it.}
\label{EV-EP-4D}
\end{figure}

\begin{figure}[bthp]
\centerline{\psfig{file=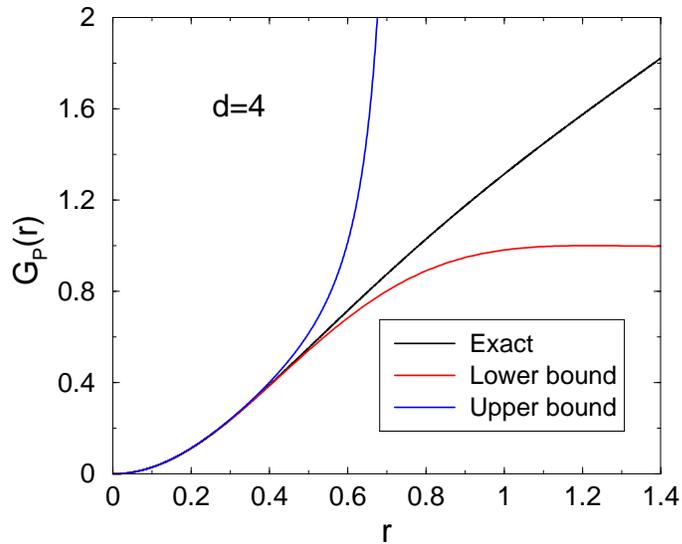,width=3.5in}}
\caption{ Upper and lower bounds on $G_P(r)$ for $d=4$ and $\rho=1$
as obtained from (\ref{EP-bound1}) and (\ref{EP-GP-bound}), respectively, compared to the corresponding 
exact evaluation of it.}
\label{GP-4D}
\end{figure}

\begin{figure}[bthp]
\centerline{\psfig{file=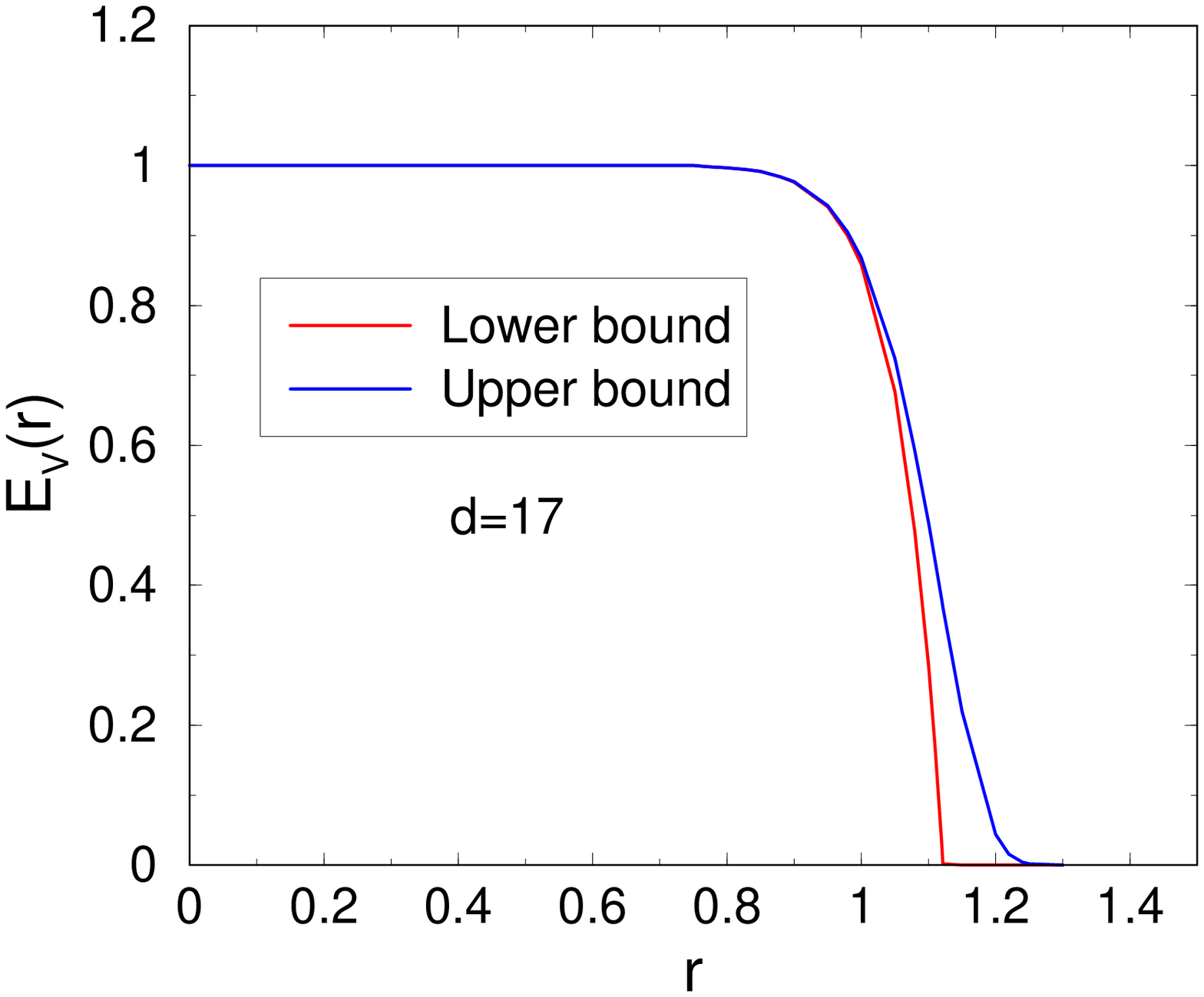,width=3.3in}\hspace{0.25in}\psfig{file=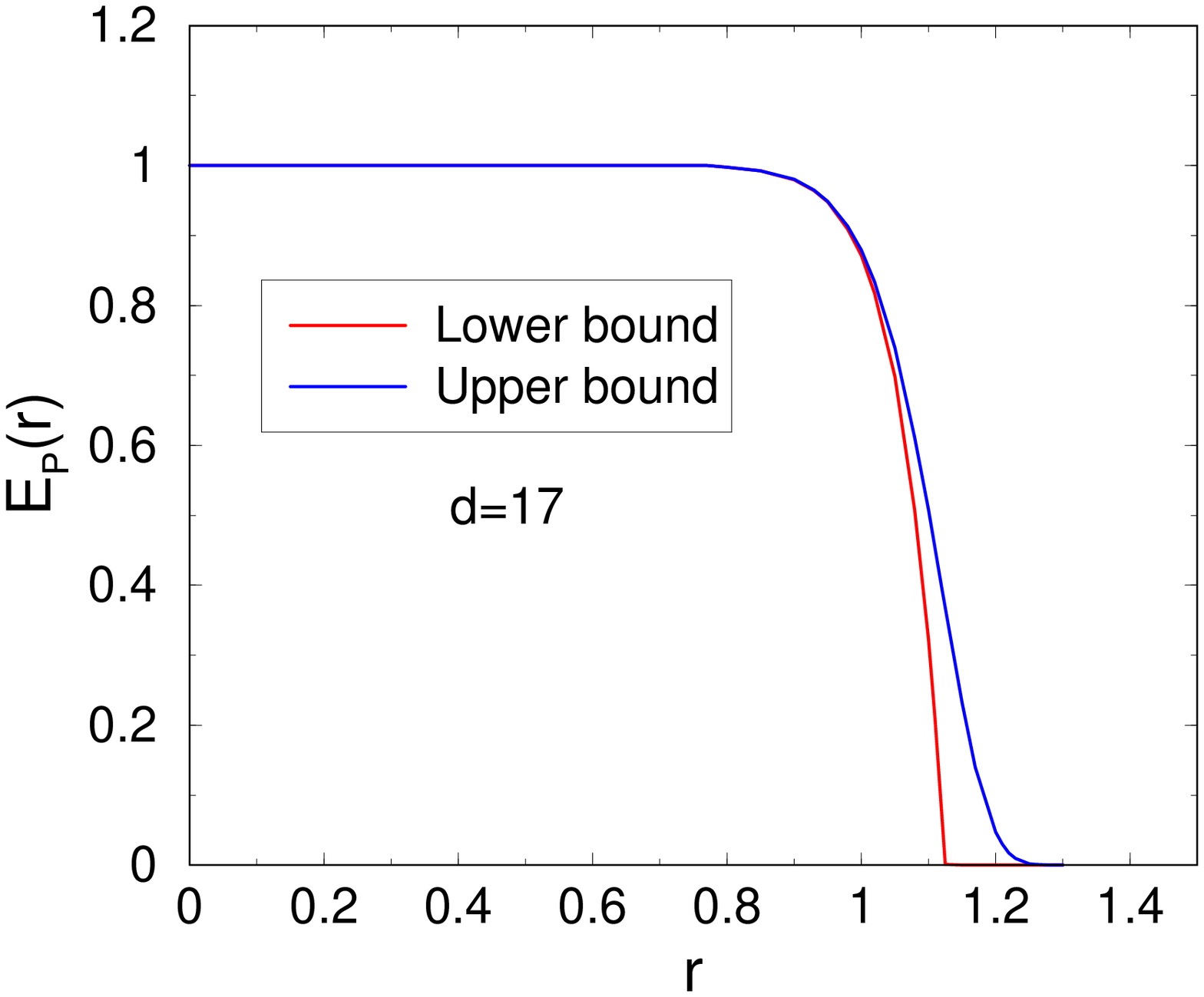,width=3.3in}}
\caption{Left panel: Upper and lower bounds on $E_V(r)$ for $d=17$ and $\rho=1$
as obtained from (\ref{EV-bound1}) and (\ref{EV-GV-bound}).
Right panel: Upper and lower bounds on $E_P(r)$ for $d=17$ and $\rho=1$
as obtained from (\ref{EP-bound1}) and (\ref{EP-GP-bound}).}
\label{EV-EP-17D}
\end{figure}

The left panel of Fig. \ref{EV-EP-4D} compares the upper bound (\ref{EV-GV-bound})
and lower bound (\ref{EV-bound1}) of $E_V$ for a Fermi-sphere point process
for $d=4$ to the corresponding exact evaluation. The right panel 
of the same figure compares the upper and lower bounds on $E_P(r)$ for $d=4$
as obtained from (\ref{EP-bound1}) and (\ref{EP-GP-bound}), respectively, to the corresponding exact evaluation of it. The upper bounds on both
exclusion probabilities provide reasonable estimates of
the exact results as compared to the corresponding lower bounds.
Figure \ref{GP-4D} compares upper and lower bounds on $G_P(r)$ for $d=4$
as obtained from (\ref{EP-bound1}) and (\ref{EP-GP-bound}), respectively, 
to the corresponding  exact evaluation of it.
Not surprisingly, the bounds become better estimators
as the dimension increases and therefore can be profitably
used in high dimensions, where exact evaluations are difficult
to obtain.
The left panel of Fig. \ref{EV-EP-17D} shows the upper bound (\ref{EV-GV-bound})
and lower bound (\ref{EV-bound1}) for a Fermi-sphere point process
for $d=17$. It is seen that the bounds essentially converge to unity
for the range $0 \le r \le 0.8$ and are relatively close
to one another for $0.8 \le r \le 1.1$. The right 
panel of Fig. \ref{EV-EP-17D} depicts
the analogous bounds on $E_P(r)$ for $d=17$. Note 
that the bounds on $E_V(r)$ behave almost exactly
like the bounds on $E_P(r)$ at this value of $d$.
This graphically suggests 
that as $d$ becomes large, the exact expressions for $E_V(r)$
and $E_P(r)$ approach the same step function,
which we demonstrate below.

%\begin{figure}[bthp]
%\centerline{\psfig{file=GV-4D-Bounds.eps,width=3.5in}\hspace{0.25in}\psfig{file=GP-4D-%Bounds.eps,width=3.5in}}
%\caption{Left panel: Upper and lower bounds on $G_V(r)$ for $d=4$
%as obtained from (\ref{GV-bound1}) and (\ref{GV-bound2})
%compared to the corresponding exact evaluation of it. 
%Right panel: Upper and lower bounds on $G_P(r)$ for $d=4$
%as obtained from (\ref{EP-bound1}) and (\ref{EP-GP-bound}) compared to the corresponding exact evaluation of %it.}
%\label{GV-GP-4D}
%\end{figure}

We now show that both $E_P$ and $E_V$ tend to the same step function $\Theta(D-r)$
in the large-$d$ limit, where  $D$ is the estimate of
the effective hard-core diameter defined by (\ref{fermihole}).
We begin by utilizing the following
upper bound on $E_P(r)$:
\begin{eqnarray}
%E_P(r) \le \Bigg\{{1, \qquad\qquad\qquad r<D,\atop{\exp[1-v_1(r)], \quad r \ge D.}}
E_P(r) \le \left\{
\begin{array}{lr}
1, \quad r<D,\\
\exp[1-v_1(r)], \quad r\ge D.
\end{array}\right.
\label{EP-bound3}
\end{eqnarray}
This upper bound is obtained by combining lower bound (\ref{Z-bounds}) on 
the cumulative coordination number $Z(r)$
and upper bound (\ref{EP-GP-bound}), and hence is a weaker upper bound on
$E_P(r)$ than (\ref{EP-GP-bound}). As $d$ tends to infinity, we see that the
upper bound (\ref{EP-bound3}) tends to the step function $\Theta(D-r)$, i.e.,
\begin{eqnarray}
E_P(r) \le \Theta(D-r) \qquad (d \rightarrow \infty).
\label{EP-upper}
\end{eqnarray}
Now we show that $E_P$ is bounded from below by the same
step function in this limit, i.e.,
\begin{eqnarray}
E_P(r) \ge \Theta(D-r) \qquad (d \rightarrow \infty).
\label{EP-lower}
\end{eqnarray}
To prove this, we note that the the lower  bound (\ref{EP-bound1}) on $E_P(r)$ tends to a unit 
step function as $d$ becomes large [cf. Fig. \ref{EV-EP-17D}] whose discontinuity location cannot 
exceed the zero of $1- Z(r)$, $D_0$, which can be estimated to be given by
\begin{eqnarray}
D_0= \left[\frac{(d+2)^2 \Gamma(1+d/2)}{d K^2}\right]^{1/(d+2)}.
\end{eqnarray}
This estimate, which bounds the zero from below and becomes increasingly accurate as 
$d$ tends to infinity, is obtained by substituting the leading term of the asymptotic expansion
(\ref{Z-small}) into $1-Z(r)$ and solving for $D_0$. 
For large $d$, $D_0$ has the asymptotic expansion
\begin{eqnarray}
D_0 = \sqrt{\frac{d}{{2\pi e }}} \left[1- \frac{\ln(d)}{2} +{\cal O}(1) \right] \qquad (d \rightarrow \infty),
\label{D0-asym}
\end{eqnarray}
Comparison of this expansion to that of the corresponding one
for $D$ [cf. (\ref{D-asym})] reveals that $D$ bounds $D_0$
from above for sufficiently large $d$ and $D_0 \rightarrow D$
in the limit $d \rightarrow \infty$.
Thus, the lower bound on $E_P(r)$ in this limit
is given by (\ref{EP-lower}). Combination of this lower
bound with upper bound (\ref{EP-upper}) leads
to the following high-dimensional behavior: 
\begin{eqnarray}
E_P(r) \rightarrow \Theta(D-r) \qquad (d \rightarrow \infty),
\end{eqnarray}
where we recall that $D$ grows like $\sqrt{d}$ for large $d$ [cf. (\ref{D-asym})].
Following the analogous analysis using the lower
bound (\ref{EV-bound1}) and upper bound (\ref{EV-GV-bound}) on $E_V(r)$, we can show 
\begin{eqnarray}
E_V(r) \rightarrow \Theta(D-r) \qquad (d \rightarrow \infty).
\end{eqnarray}

It is noteworthy that this analysis means that the lower bound (\ref{Z-bounds})
on the cumulative coordination number $Z(r)$ becomes exact
in the limit $d \rightarrow \infty$. This in turn implies an
``effective" pair correlation function $g_2^{*}(r)$ that tends to the following step 
function as $d$ tends to infinity:
\begin{eqnarray}\label{g2eff}
g_2^{*}(r) \rightarrow \Theta(r-D) \qquad (d \rightarrow \infty).
\end{eqnarray}
This effective pair correlation function $g_2^*(r)$ is to be distinguished
from the true pair correlation function (\ref{g2-d}), which tends
to unity for distances beyond the length scale $D$, but remains
a quadratic function of $r$ for small $r$. Because the effective
pair correlation function is based on the behavior of
$Z(r)$, which weights the true $g_2(r)$ by $r^{d-1}$ (due to the 
appearance of $s_1(r)$, the surface
area of a sphere of radius $r$), $g_2^*(r)$ tends to a step function
in the high-dimensional limit.
The fact that the oscillations of $g_2(r)$ seen in low dimensions 
(cf. Fig. \ref{fermions}) effectively vanish in the large-$d$ limit is consistent
with the decorrelation principle \cite{To06},  which, roughly speaking,
states that unconstrained correlations that exist
in low dimensions vanish as $d$ tends to infinity, and all higher-order correlation 
functions $g_n$ for $n\geq 3$ may be expressed in terms of $g_2$ within some small error.

We have already shown in \eref{gng2} that the latter claim is true for \emph{any}
determinantal point process.  The former claim is seen from
the form of $g_2^*$ in \eref{g2eff}, which
immediately suggests that asymptotic unconstrained correlations in the Fermi sphere point process
diminish with respect to increasing dimension $d$. 
In other words, $g_2$ flattens at unity for sufficiently large $r$ as $d$ becomes large,
which implies that long-range correlations between any two particles in the system
diminish with respect to increasing dimension, leaving only the small-$r$ correlations,
which extend outward for a greater range in $r$ as $d$ increases.  
 This conclusion in conjunction with \eref{gng2}
implies that for sufficiently large $d$ and for large particle separations, $g_n \approx \det\mathbb{I} = 1,$ where $\mathbb{I}$ is the 
$n\times n$ identity matrix.  
Therefore, all $n$-particle correlations also diminish for large particle separations and large $d$
in accordance with a decorrelation of the system.
We remark that the fact that $g_2(r) \rightarrow 0$ as $r\rightarrow 0$ does not affect the
statement of the decorrelation principle for the Fermi sphere point process; borrowing the
language of quantum mechanics, these correlations arise from the constraint of antisymmetry in
the many-particle wavefunction and therefore must be enforced in any dimension.  

For 
sufficiently large $d$ this analysis implies that the system reduces to a sphere
packing with an effective hard-core diameter equal to $D$.   
The connection to sphere packings implies that the fraction
of space $\phi$ covered by the spheres at unit number density is 
bounded from above by the following inequality
\begin{eqnarray}
\phi \le v_1(D/2) =\frac{1}{2^d}.
\end{eqnarray}
Interestingly, Minkowski proved a lower bound on the coverage fraction of the densest lattice sphere packings
that asymptotically is controlled by $1/2^d$ \cite{Mi05}.
We remark on the significance of this result in Section 6.

%, meaning that we may also view these small-$r$ correlations as constrained 
%correlations from the effective hard-core diameter.  In this light the decorrelation argument
%reduces to the one given by Torquato and Stillinger for equilibrium hard sphere systems in \cite{To06}.

%The effective $g_2$ in \eref{g2eff} allows us to define effective $n$-particle correlation functions
%$g_n^*$ in the limit $d\rightarrow +\infty$, and it is with reference to these functions that we 
%speak of a decorrelation principle via \eref{gng2}.

%The analysis immediately above demonstrates that the Fermi-sphere point process
%becomes a sphere packing in the high-dimensional limit with an effective
%hard-core diameter equal to $D$. Thus, the fraction
%of space $\phi$ covered by the spheres at unit number density is 
%bounded from above by the following inequality
%\begin{eqnarray}
%\phi \le v_1(D/2) =\frac{1}{2^d}.
%\end{eqnarray}
%Interestingly, Minkowski proved a lower bound on the coverage fraction of the densest lattice sphere packings
%that asymptotically is controlled by $1/2^d$ \cite{Mi05}.
%We remark on the significance of this result in Section VI.

\subsubsection{Mean Nearest-Neighbor Distance}

We now obtain analytical estimates of the mean nearest-neighbor distance
$\lambda$ at unit density using the general upper
and lower bounds on $\lambda$ [cf. (\ref{L1}) and (\ref{L2})]. 
Consider the arithmetic average of (\ref{L1}) and (\ref{L2}):
\begin{eqnarray}
{\overline \lambda} = \frac{\lambda_L+\lambda_U}{2}.
\end{eqnarray}
For low dimensions, the arithmetic average of the upper and
lower bounds for the Fermi-sphere point process provide
reasonable estimates of $\lambda$, as seen in Table  \ref{mean-table}
for the first four space dimensions,
which also includes the corresponding exact results.
We see that the estimate  ${\overline \lambda}$ captures
the nonmonotone dependence of the mean nearest-neighbor distance
with dimension. Moreover, the table shows that the upper bound 
becomes the dominant contribution to ${\overline\lambda}$ 
as $d$ increases and ${\overline \lambda}$ becomes increasingly
accurate as the space dimensions increases.

\begin{table}[!htp]
\caption{Comparison of estimates of the mean-nearest neighbor distance $\lambda$
for the first four space dimensions of the Fermi-sphere  point
process at unit density to the corresponding ``exact"
values. }
\begin{indented}
\item[]\begin{tabular}{c c c c c c }
\br
$d$& Upper Bound&Lower Bound & Average of Bounds& Exact & $D$\\
\mr
1 & 0.917808  &0.658199  &0.788003&   0.725728&  0.5\\  
2 & 0.688071  &0.581193  &0.634632&   0.649823&  0.564190\\  
3 & 0.670304  &0.593975  &0.632139&   0.654511&  0.620350\\
4 & 0.687631  &0.625009  &0.656320&   0.679561&  0.670938\\
\br
\end{tabular}
\label{mean-table}
\end{indented}
\end{table}

Note use of the upper bound (\ref{EP-bound3}) enables us to obtain the following weaker 
but analytically solvable upper
bound on $\lambda$:
\begin{eqnarray}
\lambda \le \lambda_U\le  \lambda_{U*}&= D+ \int_D^{\infty} \exp[1-v_1(r)] \rmd r \nonumber\\
&= D\left[ 1+\frac{\Gamma(1/d,1) e}{d}\right],
\label{lambda-bound}
\end{eqnarray}
where $\Gamma(x,a)$ is the incomplete gamma function. For large $d$,
we have the asymptotic expression
\begin{eqnarray}
\lambda_{U*} &= D\left[1 + \frac{\Gamma(0,1)\,e}{d} + {\cal O}\left(\frac{1}{d^2}\right)\right] \nonumber \\
&= D\left[1 + \frac{0.5963473622\ldots}{d} + {\cal O}\left(\frac{1}{d^2}\right)\right] \qquad (d \rightarrow \infty).
\end{eqnarray}
Moreover, using the lower bound 
\begin{eqnarray}
\lambda \ge  \int_0^{D_0} [1-Z(r)] \rmd r 
\label{lambda-bound2}
\end{eqnarray}
where $D_0$ is the zero of $1-Z(r)$, and the aforementioned asymptotic analysis
of the lower bound on $E_P(r)$, yields
\begin{eqnarray}
\lambda \ge D_0  \qquad (d \rightarrow \infty).
\end{eqnarray}

In summary, combination of the bounds
(\ref{lambda-bound}) and (\ref{lambda-bound2}) and the asymptotic expression (\ref{D0-asym}),
enables us to conclude that the mean nearest-neighbor distance approaches the length scale $D$
as $d$ becomes large, i.e.,
\begin{eqnarray}
\lambda \rightarrow D  \qquad (d \rightarrow \infty),
\label{lambda-asym}
\end{eqnarray}
which asymptotically grows like the
square root of $d$ according to (\ref{D-asym}) and, as we concluded above,
specifies the location of the step-function discontinuity
of $E_P(r)$, $E_V(r)$  and $g^{*}_2(r)$ in the large-$d$ limit. Table \ref{mean-table}
shows that the length scale $D$ is already an accurate estimate
of the mean nearest-neighbor distance for $d=4$.

It is noteworthy that the asymptotic mean nearest-neighbor-distance 
formula (\ref{lambda-asym}) 
is precisely the same as the asymptotic form of the mean nearest-neighbor
distance of a Poisson point process. The latter for any dimension
at unit density is given by $\Gamma(1+1/d)\,\Gamma(1+d/2)^{1/d}/\sqrt{\pi}$
\cite{To02a}, which in the high-dimensional limit is exactly equal to $D$. 
The fact that the mean nearest-neighbor distance
for the Fermi-sphere point process behaves like that of a Poisson
point process in the high-dimensional limit is not surprising in light of
the decorrelation principle \cite{To06}.

\begin{figure}[bthp]
\centerline{\psfig{file=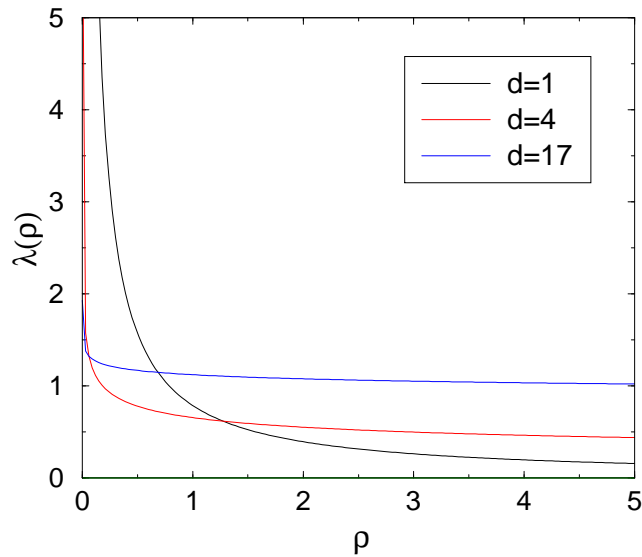,width=3.3in}}
\caption{The mean nearest-neighbor distance $\lambda$ as a function
of density $\rho$ for various dimensions. The cases $d=1$ and $d=4$ are
exact evaluations and the instance $d=17$ is obtained
from the upper bound prediction (\ref{lambda-bound}) in conjunction
with (\ref{scaling}).}
\label{mean}
\end{figure}

Note that the expression for the mean nearest-neighbor
distance $\lambda(\rho)$ for any density $\rho$ can 
be related to the corresponding quantity $\lambda(1)$
at unit density by the simple scaling relation
\begin{eqnarray}
\lambda(\rho) = \frac{\lambda(1)}{\rho^{1/d}}.
\label{scaling}
\end{eqnarray}
Figure \ref{mean} shows the mean nearest-neighbor distance as a function
of density for various dimensions; the cases $d=1$ and $d=4$ are
exact evaluations and the instance $d=17$ is obtained
from the upper bound prediction (\ref{lambda-bound}) and the
scaling relation (\ref{scaling}), which is
expected to be a highly accurate estimate.

\subsubsection{Large-$r$ Behavior}

\begin{figure}[bthp]
\centerline{\psfig{file=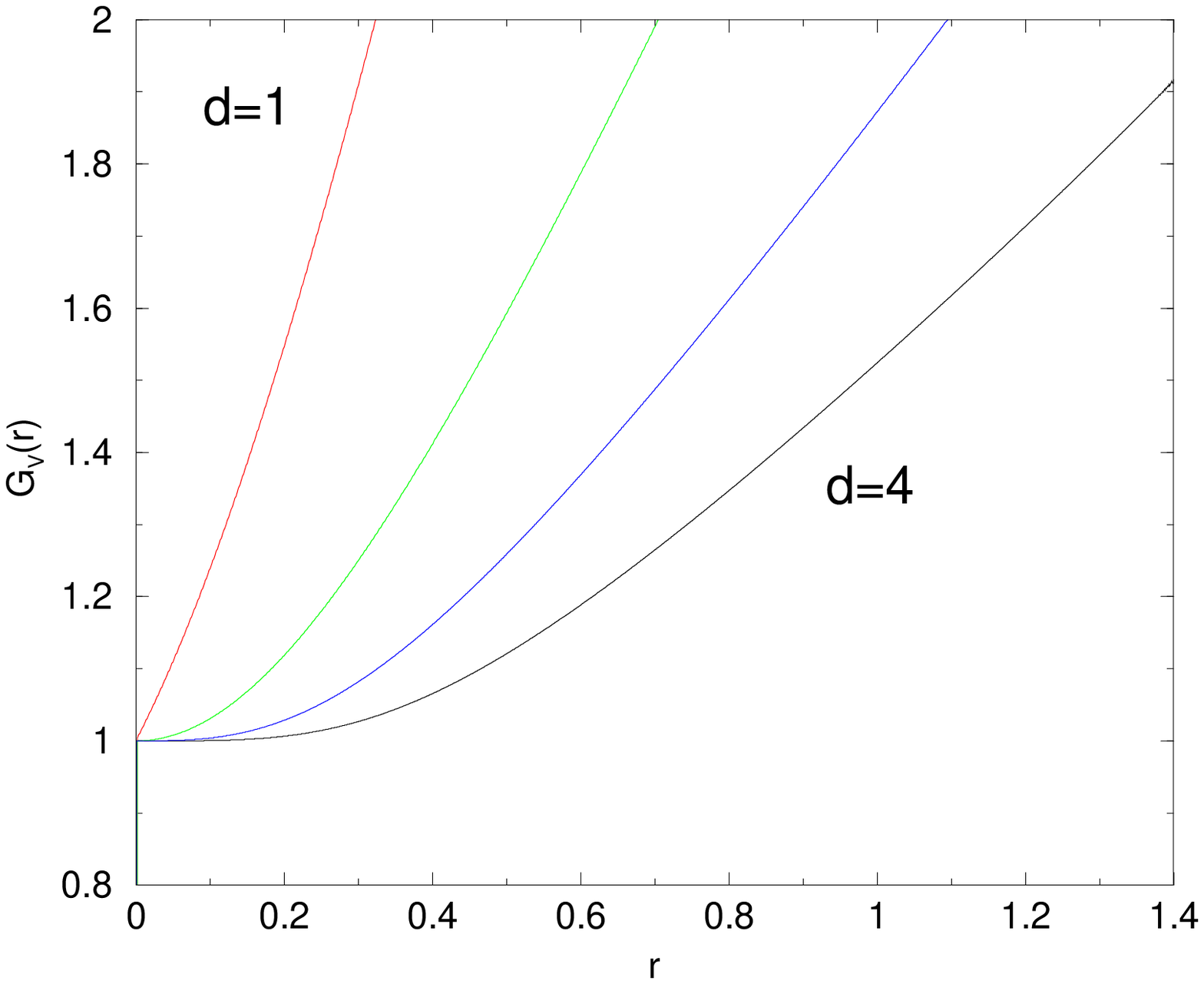,width=3.3in}\hspace{0.25in}\psfig{file=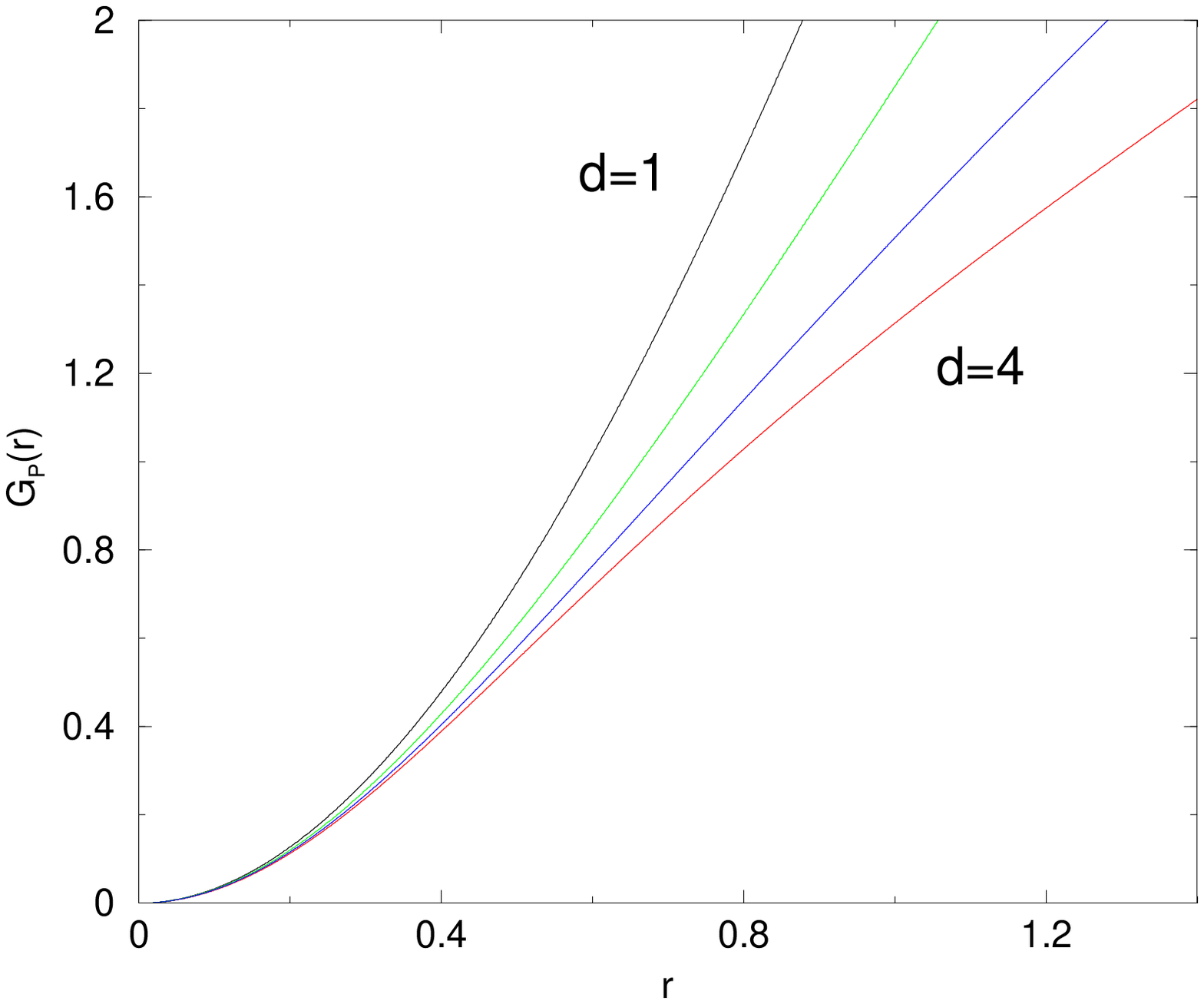,width=3.3in}}
\caption{Exact calculations of both $G_V(r)$ and $G_P(r)$
for the first four space dimensions. }
\label{GV-GP-1D-4D}
\end{figure}

We conclude this section by making some remarks about the conditional
nearest-neighbor functions $G_V(r)$ and $G_P(r)$. The fact that
the exclusion probabilities $E_V(r)$ and $E_P(r)$ tend to the
same step function in the high-dimensional limit implies
that $G_V(r)$ and $G_P(r)$ have the same large-$r$ behavior
as $d$ tends to infinity. In fact, our exact evaluations
of $G_V(r)$ and $G_P(r)$ for a finite range of $r$ 
in low dimensions indicate that each function becomes
linear in $r$ for large $r$ and the ratio $G_P(r)/G_V(r)$
tends to unity. Figure \ref{GV-GP-1D-4D} shows
our evaluations of both $G_V(r)$ and $G_P(r)$
for the first four space dimensions for the range
$0 \le r \le 1.4$ (see our companion paper \cite{Sc08}
for further numerical details).

In one dimension, one can show that the nearest-neighbor void functions
are directly related to a radial function that is a solution
to a second-order nonlinear differential equation \cite{Tr94}.
This differential equation  can be evaluated
exactly for small and large $r$. In particular,
using this asymptotic analysis leads to the following large-$r$
behavior for $G_V(r)$ for the one-dimensional Fermi-sphere point process:
\begin{eqnarray}
G_V(r) = \frac{\pi^2}{2} r + {\cal O}(r^{-1})\qquad (r \rightarrow \infty).
\end{eqnarray}
The coefficient $\pi^2/2$  can be compared to the corresponding Wigner surmise
prediction of $16/\pi$, which is obtained via the one-dimensional relation
that exactly links $E_V(r)$ to the gap distribution function $p(z)$ \cite{Ri96},
\begin{eqnarray}
E_V(r)= \int_{2r}^{\infty} (z-2r)p(z) \rmd z,
\end{eqnarray}
and Eq. (\ref{wigner}) with $\beta=2$. 
Presumably, $G_P(r)$ has the same asymptotic behavior for $d=1$,
a conclusion supported by our numerical results \cite{Sc08}.
Interestingly, for both conditional probability functions
to have the same linear behavior for large $r$ implies
that the corresponding exclusion probability functions
have the following large-$r$ behavior for finite $d$:
\begin{eqnarray}
E_V(r)=E_P(r) \rightarrow \exp[-\kappa(d) r^{d+1}] \qquad (r \rightarrow \infty),
\label{EV-poisson}
\end{eqnarray}
where $\kappa(d)$ is a positive $d$-dependent constant. 

Thus, this analysis reveals 
that the probability of finding a large spherical cavity of radius $r$ 
for a Fermi-sphere point process in dimension $d$ behaves
similar to that of a Poisson point process but in dimension
$d+1$. For a Poisson point process, the constant $\kappa(d)=\pi^{(d+1)/2}/\Gamma[(d+3)/2]$,
[cf. (\ref{E-poisson}) and therefore if $E_V(r)$ for a Fermi-sphere point process in $\mathbb{R}^d$ for large $r$
behaved exactly like that for a Poisson point process in $\mathbb{R}^{d+1}$, then $G_V(r)$ 
would be given by
\begin{eqnarray}
G_V(r) = G_P(r)=\frac{\sqrt{\pi}\,\Gamma(d/2)}{\Gamma[(d+1)/2]} r + {\cal O}(1)\qquad (r \rightarrow \infty).
\label{GV-poisson}
\end{eqnarray}
For the first four space dimensions, the coefficients multiplying
$r$ in (\ref{GV-poisson}) are given by $\pi =3.14\ldots$, 2, $\pi/2=1.57\ldots$,
and $4/3=1.33\ldots$, respectively. This can be compared to the exact
coefficients, which for $d=1,2,3$ and 4 are given by $\pi^2/2=4.93\ldots$,
$2.815\pm0.048 , 1.597 \pm 0.043$, and $1.297\pm 0.045$, respectively. 
The last three values are estimates
that we have obtained based upon extrapolations from the evaluations
of $G_V(r)$ (see Ref. \cite{Sc08} for further numerical details).
We see that the
 probability of finding a large spherical cavity of radius
$r$ in $\mathbb{R}^d$ is approximately the same as that for a Poisson point process
in $\mathbb{R}^{d+1}$, implying linear growth of $G_V(r)$ or $G_P(r)$ for 
large $r$. It is not unreasoanble to conclude that this
approximation becomes increasingly accurate as $d$ increases.  
Further justification for this remarkable behavior is given
in Section 6.

\section{Conclusions and Discussion}

We have obtained and characterized a new class of determinantal point processes 
in $\mathbb{R}^d$, the most general of which we call Fermi-shells point processes.
The $n$-particle correlation functions for any $n$ and $d$, which completely characterize
the point process, are determined analytically.
We focused primarily on a special case, the Fermi-sphere point process,
which in one dimension is identical to the point process
that characterizes the spacings of the eigenvalues of the GUE
(as well as the CUE), the conjectured spacings of the nontrivial
zeros of the Riemann zeta function, and the positions of
spin-polarized fermions in their ground state (i.e., completely filling the Fermi ``sphere").
We are not aware of any correspondence of the general
Fermi-shells point process in $\mathbb{R}^d$ for any $d \ge 2$ to random matrix theory
or the zeros of any generalized zeta function in number
theory.   If our determinantal point processes have connections to random matrices
in arbitrary space dimension, the latter must be non-Hermitian. For example, Ginibre
\cite{Gi65} showed that there are two-dimensional determinantal point processes
that correspond to complex eigenvalues of random  non-Hermitian matrices. Thus, it is an open question
whether the Fermi-shells point process
has any correspondence to random matrices.

We analyzed in great detail properties of  pair statistics, including the pair correlation
function, the structure factor, and cumulative coordination number,
as a function of spatial dimension $d$. The point processes
for any $d$ are shown to be hyperuniform such that the structure factor (or power spectrum) $S(k)$ has
a nonanalytic behavior at the origin given by $S(k) \sim |k|$ ($k \rightarrow 0$).
The latter result implies that the pair correlation $g_2(r)$ 
tends to unity  for large pair distances with a decay rate
that is controlled by the power law $1/r^{d+1}$. In three dimensions,
such a dominant power-law decay of $g_2(r)$ is a well-known
property of bosonic systems in their ground states \cite{Fe54,Re67} and, more
recently, has been shown to characterize  maximally random jammed sphere packings \cite{Do05}.
We also graphically displayed one- and two-dimensional realizations 
of the point processes in order to vividly reveal
their ``repulsive" nature and demonstrated that they
can be characterized by an effective ``hard-core" diameter
that grows like the square root of $d$.

Our study of the nearest-neighbor functions of the Fermi-sphere point
process resulted in some noteworthy conclusions. For example, 
we have seen that the probability of finding a large spherical cavity of radius
$r$ in $\mathbb{R}^d$ is approximately the same as that for a Poisson point process
in $\mathbb{R}^{d+1}$, implying linear growth of $G_V(r)$ or $G_P(r)$ for 
large $r$. This is a remarkable result because it represents
the first class of nontrivial point processes that we are aware of whose conditional
nearest-neighbor functions do not asymptote to a constant; see Refs. \cite{To90,To90b,Lu92}
and \cite{Ri96} for examples of correlated equilibrium and nonequilibrium point
processes that possess $G_V(r)$ or $G_P(r)$ with constant asymptotes, respectively. 
It is clear that the unusual property 
of linear growth of $G_V$ or $G_P$ for large $r$ is due to the  long-range nature of the 
repulsive interactions. Since a Fermi-sphere point process in $\mathbb{R}^d$ is always hyperuniform, i.e.,
large wavelength density fluctuations vanish, the probability of finding a
large spherical cavity must be smaller than the corresponding quantity
for any point process that is not hyperuniform, such as a Poisson point
process in $\mathbb{R}^d$. The probability of finding a large spherical
cavity is of course smaller in a Poisson point process in dimension $d+1$
compared to that in dimension $d$.  Moreover, it is easy to
show that the asymptotic form  (\ref{EV-poisson}) for $E_V(r)$ or $E_P(r)$ for large $r$ is 
always between the aforementioned corresponding rigorous upper and lower bounds 
on the exclusion probabilities.

We also found that the Fermi-sphere point process
becomes a sphere packing in the high-dimensional limit with an effective
hard-core diameter equal to the length scale $D$ [cf. (\ref{fermihole})]. Thus, the fraction
of space $\phi$  covered by the spheres at unit number density is 
bounded from above by $1/2^d$. This coverage fraction has a special
significance in the study of sphere packings; it arises
not only in Minkowski's  famous century-old lower bound 
on the density of the densest lattice sphere packings \cite{Mi05}
but  in lower bounds for saturated and disordered packings \cite{To06}
as well as the highest achievable density in the ``ghost" random
sequential addition packing \cite{To06b}. It should be noted,
however, that there is strong evidence that there exist disordered sphere
packings with $\phi$ not only greater than $1/2^d$ \cite{Pa06,Sk06,To06c,Za07}
but with densities that exponentially improve
on Minkowski's lower bound \cite{To06,Sc08b}.

Elsewhere \cite{Sc08} we report results on the extremes of the nearest-neighbor
statistics as well as Voronoi statistics of the Fermi-sphere point
processes in the first four space dimensions. In other work, we will
quantify clustering and percolation properties of Fermi-sphere point
processes.

\ack
The authors are grateful to Peter Sarnak and Juan Maldacena
for useful discussions.
This work was supported by the National Science Foundation
under Grant No. DMS-0804431.

\appendix

%\section{Filling the Fermi Sphere and Interaction Potential}
\section{On the Presence of Intrinsic $n$-Particle Interactions
(with $n \ge 3$) for General Determinantal Point Processes}

Our purpose here is to show that the $n$-particle probability density
function for an arbitrary determinantal point process, even in one
dimension, cannot be written as a Boltzmann factor of $N$ classical particles
interacting through one- and two-body potentials at a finite temperature.
Although this claim is true for each of the canonical ensembles of 
random-matrix theory, we show via a counterexample in one dimension 
that intrinsic $n$-particle interactions with
$n \ge 3$ are generally necessary to describe the system.  

It is relatively straightforward to express the $n$-particle probability
density function for the CUE as a Boltzmann factor of a classical system of
pairwise-interacting particles. In analogy with the
formalism of the one-dimensional Fermi-sphere point process
introduced in Section 4.1, we fill the Fermi  
line at a constant density. Thus, may write \cite{Me67}:
\begin{eqnarray}
\det[\exp(\rmi n x_m)]_{n,m}=\prod_{n<m}[\exp(\rmi x_n)-\exp(\rmi x_m)];
\end{eqnarray}
therefore,
\begin{eqnarray}
\left|\det[\exp(\rmi n x_m)]_{n,m}\right|^2&=\exp\left[-2\sum_{n<m}\ln|\exp(\rmi x_n)- 
\exp(\rmi x_m)|\right]\\
&=\exp\left[-2\sum_{n<m}\ln|\sin(x_n-x_m)/2|\right]\label{A3}.
\end{eqnarray}
For small eigenvalue separations $\Delta x_{nm} = x_n - x_m$, the result in 
\eref{A3} reduces to the form:
\begin{eqnarray}
|\det[\exp(\rmi nx_m)]_{n,m}|^2 \approx \exp\left[-2\sum_{n<m}\ln|(\Delta x_{nm}/2)|\right],
\end{eqnarray}
and the probability distribution of the CUE eigenvalues
can indeed be written as a pairwise interacting  
potential, which in this example is logarithmic for small
particle separations. It is important to note that this reformulation of the probability density
is very peculiar to the method of  
filling the Fermi sphere. Suppose instead that we decide to fill only the  
states $n=0,2,3, \ldots$; i.e., the state $n=1$ is skipped. We may then express the three-body form of the probability density as:
\begin{eqnarray}
\fl \left|\det[\exp(\rmi n x_m)]_{n,m}\right|^2=\frac{64}{(2\pi)^3} [3+2  
\cos(x-y)+2 \cos(x-z)+2 \cos(y-z)]\nonumber\\ 
\times 
\sin\left(\frac{x-y}{2}\right)^2 \sin\left(\frac{x-z}{2}\right)^2  
\sin\left(\frac{y-z}{2}\right)^2,
\end{eqnarray}
where we have used $x, y,$ and $z$ to represent $x_0, x_2,$ and $x_3$; the subscripts denote the state $n$ of each particle.
The last three factors have the form of the pair interaction in \eref{A3}. 
To check whether the pre-factor containing cosines can  
be written in the same form, we write:
\begin{eqnarray}
V(x,y,z)=-\ln[3+2 \cos(x-y)+2 \cos(x-z)+2 \cos(y-z)]\label{threepot}.
\end{eqnarray}
Assume that $V(x,y,z)=v(x,y)+v(y,z)+v(z,x)$. It must then be true that:
\begin{eqnarray}
v(x,y)&=\frac{1}{2}\left[V(x,y,y)-\frac{1}{3}V(y,y,y)\right]\\
&=\frac{\ln(3)}{3}
+\frac{1}{2} \ln[5+4 \cos(x-y)],
\end{eqnarray}
but by substituting this expression into \eref{threepot}, we see that
we do not recover the original  
functional form. Therefore, we have shown by contradiction that for general determinantal point 
processes, the interaction potential must contain at least intrinsic three-body terms.


\section*{References}

\begin{thebibliography}{20}




\bibitem{Dy62}
Dyson F J, \emph{Statistical Theory of the Energy Levels of Complex Systems. I}, 
1962 \emph{J. Math. Phys.} {\bf 3} 140

\bibitem{Dy70}
Dyson F J, \emph{Correlations between Eigenvalues of a Random Matrix},
1970 \emph{Comm. Math. Phys.} {\bf 19} 235

\bibitem{Me67}
Mehta M L, 1967 {\it Random Matrices} (New York:  Academic Press)

\bibitem{Mo73}
Montgomery H L, \emph{The pair correlation of zeros of the zeta function}, in: 
1973 \emph{Proc. Sympos. Pure Math.}
vol. 24 (Providence: AMS) pp. 181-193

\bibitem{Od87}
Odlyzko A M, \emph{On the Distribution of Spacings Between Zeros of the Zeta Functions},
1987 \emph{Math. Comput.} {\bf 48} 273;
Odlyzko A M, \emph{The $10^{22}$-nd zero of the Riemann zeta function}, in:
van Frankenhuysen M and Lapidus M L, 2001 \emph{Dynamical, Spectral, and Arithmetic Zeta Functions}
(Providence:  AMS) pp. 139-144

\bibitem{Ru96}
Rudnick Z and Sarnak P, \emph{Zeros of principal $L$-functions and random matrix theory},
1996 \emph{Duke Math. J.} {\bf 81} 269

\bibitem{Ka99}
Katz N M and Sarnak P, \emph{Zeros of zeta functions and symmetry}, 
1999 \emph{Bull. Amer. Math. Soc.} {\bf 36} 1

\bibitem{To06b}
Torquato S and Stillinger F H, \emph{Exactly solvable disordered sphere-packing model in arbitrary-dimensional Euclidean spaces}, 
2006 \emph{Phys. Rev. E} {\bf 73} 031106

\bibitem{Fe98}
Feynman R P, 1998 {\it Statistical Mechanics} (Boulder, CO:  Westview Press)

\bibitem{Fe54}
Feynman R P, \emph{Atomic Theory of the Two-Fluid Model of Liquid Helium},
1954 \emph{Phys. Rev.} {\bf 94} 262

\bibitem{Fe56}
Feynman R P and Cohen M, \emph{Energy Spectrum of the Excitations in Liquid Helium},
1956 \emph{Phys. Rev.} {\bf 102} 1189

\bibitem{Re67}
Reatto L and Chester G V, \emph{Phonons and the Properties of a Bose System},
1967 \emph{Phys. Rev.} {\bf 155} 88

%\bibitem{Pe93}
%P. J. E. Peebles, {\it Principles of Physical Cosmology}
%(Princeton University Press, Princeton, New Jersey, 1993).


\bibitem{Do05}
Donev A, Stillinger F H and Torquato S, \emph{Unexpected Density Fluctuations in Jammed Disordered Sphere Packings},
2005 \emph{Phys. Rev. Lett.} {\bf 95} 090604

\bibitem{Sc08}
Scardicchio A, Zachary C E and Torquato S, 2008 submitted for publication

\bibitem{St95}
Stoyan D, Kendall W S and Mecke J,
1995 {\em Stochastic Geometry and Its Applications}
(New York:  Wiley)

\bibitem{To06}
Torquato S and Stillinger F H, \emph{New Conjectural Lower Bounds on the Optimal Density of Sphere Packings},
2006 \emph{Exp. Math.} {\bf 15} 307

\bibitem{Beck87}
Beck J, \emph{Irregularties of distribution {I}},
1987 \emph{Acta Mathemtica} {\bf 159} 1

%\bibitem{Bo36}
%S.~Bochner.
%\newblock {\em Lectures on Fourier Analysis}.
%\newblock Edwards, Ann Arbor, Michigan, 1936.

\bibitem{To03a}
Torquato S and Stillinger F H, \emph{Local density fluctuations, hyperuniformity, and order metrics},
2003 \emph{Phys. Rev. E} {\bf 68} 041113

\bibitem{Ga03}
Gabrielli A, Jancovici B, Joyce M, Lebowitz J L, Pietronero L and Labini F S, \emph{Generation of primordial cosmological perturbations from statistical mechanical models},
2003 \emph{Phys. Rev. D} {\bf 67} 043506

\bibitem{Ga04}
Gabrielli A and Torquato S, \emph{Voronoi and void statistics for superhomogeneous point processes},
2004 \emph{Phys. Rev. E} {\bf 70} 041105

\bibitem{Ga08}
Gabrielli A, Joyce M and Torquato S, \emph{Tilings of space and superhomogeneous point processes},
2008 \emph{Phys. Rev. E} {\bf 77} 031125

\bibitem{To02c}
Torquato S and Stillinger F H, \emph{Controlling the Short-Range Order and Packing Densities of Many-Particle Systems}, 
2002 \emph{J. Phys. Chem. B} {\bf 106} 8354;
{\it ibid}, 2002 {\bf 106} 11406

\bibitem{Sc08b}
Scardicchio A, Stillinger F H and Torquato S, \emph{Estimates of the optimal density of sphere packings in high dimensions},
2008 \emph{J. Math. Phys.} {\bf 49} 043301 

\bibitem{footnote}
A sphere packing is a point process in which there is a minimal positive
distance between any pair of points. According to Ref. \cite{To06},
a {\it disordered} sphere packing in $\mathbb{R}^d$
is one in which the pair correlation function $g_2({\bf r})$ decays
to its long-range value of unity faster than $|{\bf r}|^{-d-\varepsilon}$ for
some $\varepsilon >0$.

\bibitem{Sa06}
Sarnak P and Strombergsson A, \emph{Minima of Epstein's Zeta function and heights of flat tori},
2006 \emph{Inventiones Math.} {\bf 165} 115

\bibitem{To08}
Torquato S and Stillinger F H, \emph{New Duality Relations for Classical Ground States}, 
2008 \emph{Phys. Rev. Lett.} {\bf 100} 020602

%\bibitem{Ku07}
%Realizability of Point Processes
%T. Kuna, J. L. Lebowitz and E. R. Speer, 
%Realizability of Point Processes	
%J. Stat. Phys. {\bf 129}, 417 (2007).


%\bibitem{Ya61}
%M.~Yamada.
%\newblock Geometrical study of the pair distribution function in the many-body
%  problem.
%\newblock {\em Prog. Theor. Phys.}, 25:579--594, 1961.



\bibitem{Ma75}
Macchi O, \emph{The Coincidence Approach to Stochastic Point Processes},
1975 \emph{Adv. Appl. Probab.} {\bf 7} 83

\bibitem{So00}
Soshnikov A, \emph{Determinantal random point fields},
2000 \emph{Russian Mathematical Surveys} {\bf 55} 923

\bibitem{Ho06}
Hough J B, Krishnapur M, Peres Y and Vir{\' a}g B, \emph{Determinantal Processes and Independence},
2006 \emph{Probab. Surveys} {\bf 3} 206

\bibitem{Jo04}
Johansson K, \emph{Determinantal Processes with Number Variance Saturation},
2004 \emph{Comm. Math. Phys.} {\bf 252} 111

\bibitem{Bu93}
Burton R and Permantle R, \emph{Local Characteristics, Entropy and Limit Theorems for Spanning Trees and Domino Tilings Via Transfer-Impedances},
1993 \emph{Ann. Probab.} {\bf 21} 1329

\bibitem{HoJo05}
Horn R A and Johnson C R, 
2005 \emph{Matrix Analysis} (Cambridge:  Cambridge UP)

\bibitem{anydensity}
Note that for arbitrary density $\rho$, the corresponding
pair correaltion function is still given by (\ref{g2-d-2})
but with $K=\rho^{1/d} 2\sqrt{\pi}\,\Gamma(1+d/2)^{1/d}$,
which reduces to (\ref{R}) for unit density.

\bibitem{GiKl06}
Gioev D and Klich I, \emph{Entanglement Entropy of Fermions in Any Dimension and the Widom Conjecture},
2006 \emph{Phys. Rev. Lett.} {\bf 96} 100503

\bibitem{Cos04}
Costin O and Lebowitz J L, \emph{On the Construction of Particle Distributions with Specified Single and Pair Densities},
2004 \emph{J. Phys. Chem. B.} {\bf 108} 19614


\bibitem{To93}
Torquato S and Lu B, \emph{Chord-length distribution function for two-phase random media},
1993 \emph{Phys. Rev. E}  {\bf 47} 2950

\bibitem{To02a}
Torquato S, 2002 \emph{Random Heterogeneous Materials: Microstructure and Macroscopic Properties}
(New York:  Springer-Verlag)

\bibitem{To90}
Torquato S, Lu B and Rubinstein J, \emph{Nearest-neighbor distribution functions in many-body systems},
1990 \emph{Phys. Rev. A} {\bf 41} 2059

\bibitem{Mc60}
McWeeny R, \emph{Some Recent Advances in Density Matrix Theory},
1960 \emph{Rev. Mod. Phys.} {\bf 32} 335

\bibitem{Sl51}
Slater J C, \emph{A Simplification of the Hartree-Fock Method},
1951 \emph{Phys. Rev.} {\bf 81} 385

\bibitem{BoCo74}
Boyd R J and Coulson C A, \emph{The Fermi hole in atoms},
1974 \emph{J. Phys. B:  Atom. Molec. Phys.} {\bf 7} 1805

\bibitem{Sc05}
Schwabl F, 2005 {\it Advanced Quantum Mechanics} 3rd ed. (Berlin:  Springer-Verlag)

\bibitem{Mi05}
Minkowski H, \emph{Diskontinuit{\" a}tsbereich f{\" u}r arithmetische {\" A}quivalenz},
1905 \emph{J. reine angew. Math.} {\bf 129} 220

\bibitem{Tr94}
Tracy C and Widom H, \emph{Level-Spacing Distributions and the Airy Kernel},
1994 \emph{Commun. Math. Phys.} {\bf 159} 151

\bibitem{Ri96}
Rintoul M D, Torquato S and Tarjus G, \emph{Nearest-neighbor statistics in a one-dimensional random sequential adsorption process}, 
1996 \emph{Phys. Rev. E} {\bf 53} 450 

\bibitem{Gi65}
Ginibre J, \emph{Statistical Ensembles of Complex, Quaternion, and Real Matrices},
1965 \emph{J. Math. Phys.} {\bf 6} 440

\bibitem{To90b}
Torquato S and Lee S B, \emph{Computer simulations of nearest-neighbor distribution functions and related quantities for hard-sphere systems},
1990 \emph{Physica A} {\bf 164} 347 

\bibitem{Lu92}
Lu B and Torquato S, \emph{Nearest-surface distribution functions for polydispersed particle systems},
1992 \emph{Phys. Rev. A} {\bf 45} 5530 

\bibitem{Pa06}
Parisi G and Zamponi F, \emph{Amorphous packings of hard spheres for large space dimension},
2006 \emph{J. Stat. Mech.: Theory Exp.} P03017

\bibitem{Za07}
Zamponi F, \emph{Some recent theoretical results on amorphous packings of hard spheres},
2007 \emph{Phil. Mag.} {\bf 87} 485 

\bibitem{Sk06}
Skoge M, Donev A, Stillinger F H and Torquato S, \emph{Packing hyperspheres in high-dimensional Euclidean spaces},
2006 \emph{Phys. Rev. E} {\bf 74} 041127 

\bibitem{To06c}
Torquato S, Uche O U and Stillinger F H, \emph{Random sequential addition of hard spheres in high Euclidean dimensions},
2006 \emph{Phys. Rev. E} {\bf 74} 061308




\end{thebibliography}
\end{document}